\documentclass[12pt,a4paper]{article}

\pdfoutput=1 

\usepackage{booktabs} 
\usepackage{adjustbox}
\usepackage{amsmath}
\usepackage{amssymb}
\usepackage{url}
\usepackage[comma]{natbib}
\usepackage{appendix}
\usepackage{changepage}
\usepackage{setspace}
\usepackage{color}
\definecolor{darkblue}{rgb}{0.0, 0.0, 0.55}
\definecolor{shortengray}{gray}{0.5}
\definecolor{gray}{gray}{0.5}
\usepackage[flushleft]{threeparttable}
\usepackage{bbm}
\usepackage{lscape}
\usepackage{float}

\makeatletter
\let\orig@table\table         \let\orig@endtable\endtable
\let\orig@figure\figure       \let\orig@endfigure\endfigure
\renewenvironment{table}[1][]{\orig@table[!htbp]}{\orig@endtable}
\renewenvironment{figure}[1][]{\orig@figure[!htbp]}{\orig@endfigure}
\makeatother
\usepackage{flafter}          

\usepackage{subcaption} 
\usepackage{times}
\usepackage{ragged2e}
\usepackage{multirow,array}
\usepackage{xr}
\usepackage{rotating}
\usepackage{enumitem} 
\usepackage{tabulary}

\usepackage{pgffor} 

\usepackage{blindtext}
\usepackage{setspace}

\usepackage[comma]{natbib}
\usepackage{sectsty}
\allsectionsfont{\normalsize \bfseries}

\usepackage[compact]{titlesec}

\titleformat{\section}{\normalfont\fontsize{15}{15}\bfseries}{\thesection}{1em}{}

\titlespacing{\section}{1pt}{2ex}{1ex}
\titlespacing{\subsection}{1pt}{2ex}{1ex}
\titlespacing{\subsubsection}{1pt}{2ex}{1ex}

\usepackage[paper=a4paper,left=1in,right=1in,top=1in,bottom=1in]{geometry}

\usepackage[english]{babel}
\usepackage{subcaption} 
\usepackage{wrapfig} 
\usepackage[hidelinks]{hyperref} 

\title{\Large \textbf{Local Media and the Shaping of Social Norms: \\ Evidence from the Ebola outbreak}}

\author{\textsc{Ada Gonz\'{a}lez-Torres} \thanks{\href{mailto: adagt@bgu.ac.il}{adagt@bgu.ac.il}, Department of Economics, Ben-Gurion University of the Negev, P.O. Box 653, Be'er Sheva 8410501, Israel. I am grateful J\'{e}r\^{o}me Adda, Mich\`{e}le Belot, Edward A. Miguel and Noam Yuchtman for invaluable support and advice. I am thankful to Michael Anderson, Eliana La Ferrara, David K. Levine, Fred Finan, Peter R. Hansen, Andrea Ichino for their guidance. I am grateful to Maxim Ananyev, Darin Christensen, Brian Dillon, Ray Fisman, Thomas Fujiwara, Manoel Gehrke Ryff Moreira, Amma Panin, Dan Posner, Amanda Robinson, and Miri Stryjan for detailed feedback. I am thankful to Anjali Adukia, Jorge Ale-Chilet, Marcella Alsan,  Laura Boudreau, Ylenia Brilli, Sylvain Chassang, Itzel De Haro, Auguste Debroise, Stefano Della Vigna, Oded Galor, Jorge García Hombrados, Erick Gong, Kenneth Houngbedji, Sarah Lowes, Ben Marx, Andreas Madestam, John Marshall, Luigi Minale, Eduardo Montero, Karthik Muralidharan, Matthew Neidell, Matthew Notowidigdo, Elias Papaioannou, Vincent Pons, Ray Fisman, Claude Raisaro, Manaswini Rao, Yona Rubinstein, Raul S\'{a}nchez de la Sierra, Awa Ambra Seck, Jesse Shapiro, Itai Trilnick, and to anonymous referees for valuable feedback on this research and to seminar participants at UC Berkeley, BGU, Bar-Ilan, UCLA, the European University Institute, ETH Zurich, Hebrew University, Reichman University, as well as participants at the Society for Institutional and Organizational Economics Conference in Chicago, the Pacific Development Conference, the Working Group in African Political Economy Annual Meeting, the NCID/Fundacion Areces Workshop, the IEA Conference at Hebrew University, the Spring Meeting for Young Economists, the Arne Ryde Workshop on Culture, Institutions, and Development at Lund University, the Spanish Development Economics Workshop (SPANDE), the Econometric Society Africa meeting in Cairo, and anonymous referees. I am indebted to Ren\'{e} Migliani for facilitating contacts in Guinea and Maarten Voors for access to funding for data collection in Guinea. I thank Moustapha Diop, Regina Ellwanger, Alain Epelboin, Kenneth Houngbedji, Ibrahima Kaba, Thomas Kratz, and local support for their welcome and guidance on the medical, anthropological and local context of the epidemic, Roger Coudé, Peer-Axel Kroeske, Eric Lehmer and Günter Lorenz for advice on radio transmission software and GIS. I thank Justin Black, Sam Church and Didi Egerton-Warburton for excellent research assistance. I gratefully acknowledge funding from the Royal Netherlands Embassy in Ghana for research in Guinea, and support from the Weatherhead Center for International Affairs at Harvard University.}} 

\date{
\vspace{.5cm}
\today
}

\def\dirnew{.}                    

\def\maps{./output/maps}

\def\dirout{./output/_paper_org}
\def\figures{\dirout/figures}

\def\mapsraw{\maps/raw}
\def\mapfig{\maps/python}

\input{\dirnew/bibtex/tcilatex}

\def\bootrob{boot} 

\def\hl{h} 

\def\radio{\hl} 
\def\lang{MED} 

\makeatletter
\newcommand\medsize{\@setfontsize\medsize{14}{16}} 
\makeatother

\begin{document}
	\maketitle


\vspace*{-1cm}
\begin{abstract}
Media’s influence on norms and behavior is widely recognized. Less is known about the role played by media being local. I examine this in a high-stakes context, the Ebola outbreak in Guinea. I exploit quasi-random variation in access to radio and the timing of a public-health campaign aired on community radio. I find that 12-17\% of Ebola cases could have been prevented if places with access to a neighboring community radio station had instead had their own. Impacts are driven by radio being local rather than by ethno-linguistic belonging. Local media facilitates coordination in behaviors observed and sanctioned locally.
\end{abstract}

\thispagestyle{empty}

\clearpage

Mass media has been shown to shape a wide range of socio-economic outcomes, from public spending and voting to gender norms and violence \citep{stromberg2004radio, dellavigna2007fox, jensen2009power, ferrara2012soap, yanagizawa2014propaganda, armand2020reach}.\footnote{See \citet{stromberg2015media, dellavigna2015economic} for reviews.} Yet evidence that information changes health behavior is mixed at best \citep[e.g.][]{dupas2011teenagers, duflo2015education, galiani2016promoting, banerjee2024messages, dupas2024negligible, wakefield2010use}. Many health behaviors are tightly linked to community norms: they are observed, and often sanctioned, by the community, and enacted through ritual and ceremony, and can persist even when costly or harmful.\footnote{Religious festivals persist though they depress incomes when they coincide with the harvest \citep{montero2022religious}, and norms such as female genital cutting and child marriage survive across generations \citep{de2019does, corno2020age}.} Anthropological accounts suggest that shifting such behaviors requires communication that is interpersonal rather than simply broadcast \citep{richards2016ebola}. But interpersonal communication lacks the coordination through which media can shift norms. Can local media resolve this tension, being socially embedded while facilitating coordination? 


This paper proposes that local media can be both socially embedded and coordinating, and that this is key for some behaviors, but not others. It highlights a new trade-off between the reach of media and its influence on social norms: media with greater coverage is generally thought to shape behavior more effectively, by reaching a larger audience, but broad communication updates beliefs about many, mostly distant, others, while barely moving beliefs about the people physically close by. Local media coordinates less broadly than national or global media, but perhaps at the level that matters most for socially-entrenched behaviors: the community that observes and sanctions them. Such coordination should matter little for behaviors that are private or not reprimanded, like hand-washing, but may be central for behaviors sanctioned by the community, such as mask-wearing, burial practices, and vaccine uptake. Local media may therefore be more effective precisely for these latter behaviors—though this could equally reflect trust or relevance to local listeners rather than coordination per se. 

I study the impacts of local media on health behavior and the ultimate spread of disease, and identify the channel through which it operates, in the high-stakes context of the 2014--16 Ebola epidemic in West Africa. The outbreak killed over 2,500 people in Guinea alone and over 11,000 across West Africa, making it the deadliest Ebola epidemic in history. In response to the outbreak, Guinea launched a major public health campaign, designed and coordinated from the capital, yet aired in local languages across the country's community radio stations. The campaign aimed at convincing the population to change health behaviors attached to cultural practices and prevalent conspiracy theories that facilitated the spread of Ebola. 

Identifying the impacts of media being local is not straightforward. It requires variation in access to local media outlets, holding other things constant: informational content and access to a similar media outlet that is not local. Decoupling the effect of locality from trust or relevance due to shared ethnic group or language is also difficult: it necessitates enough variation in ethno-linguistic composition of media outlets and within geographic locations. To evaluate the role of media in coordinating behavior, we need to observe outcomes attached to social norms. 


In a difference-in-differences design, combined with novel data collected on-site, I examine the spread of Ebola in places with more or less access to a given radio station, before and after the launch of the public health campaign. This design exploits two sources of quasi-random variation. The first is given by the cross-sectional variable: access to radio before the start of the outbreak. For each radio station, I construct a measure of radio access based on the technical attributes of transmitters and the geographic features of the terrain, using the Irregular Terrain Model (ITM), developed by engineers and standard in the economics literature since \citet{olken2009television}. Guinea's terrain ruggedness generates plausibly exogenous variation in radio reception.\footnote{Guinea's elevation spans from 0 to 1752 meters above sea level. The ruggedness index developed by \citet{nunnpuga2012} shows that Guinea's ruggedness is among the top half of African countries, not as high as in Rwanda, but similar to Indonesia, and twice the ruggedness of Congo, where the ITM has also been applied to study impacts of media \citep[see][]{olken2009television, yanagizawa2014propaganda, armand2020reach}.} The second source of quasi-random variation is the timing of the public health campaign, launched several months after the start of the epidemic. As baseline, I use the official campaign announcement date (six months into the epidemic) as the reference period, rather than station-specific adoption dates, and report effects relative both to this date and to the mass scale-up three months later. This avoids endogeneity issues from radio stations gradually adopting public health messages as the epidemic spreads, and identifies lower-bound impacts; to the extent that adoption was gradual, effects may be observed with a time lag relative to the official start. I then complement this with evidence from a subset of radio stations for which I have effective adoption dates, using a dynamic difference-in-differences design \citep{de2024difference} to examine whether impacts are more immediate. The main outcomes are monthly Ebola infections, obtained from patient records from the World Health Organization (WHO) in Guinea, complemented by measures of social health behavior.



To study the role of the locality of media, I exploit the fact that the reach of radio can go beyond geographic boundaries. I can therefore differentiate between places with access to a community radio station based in their own community or region, compared to neighboring places reached by the same radio station, but for which it is not local. Further, given high levels of diversity within and across regions, I also study whether the effects of community radio are driven by the ethno-linguistic makeup of its listeners instead of locality. I compare outcomes in places with access to a community radio station with shared language, to places with access to the same radio, but with a different majority language.

The results show that the public health campaign aired on community radio led to an earlier drop and a greater reduction in Ebola cases in places with a community radio station from their own community, compared to places with access to the same radio station but for which it is not local (Section \ref{sec:empirical}). The main result is perceptible from raw data (Figure~\ref{fig:did_main}). During the initial months of the outbreak, new infections grow at a similar rate in places with and without access to a community radio station that is local to them, conditional on above-median overall access to community radio. Following the launch of the campaign, Ebola infections drop in places with a community radio that is local to them. A back-of-the-envelope calculation implies that 447-664 infections, 12-17\% of total Ebola cases in Guinea, would have been prevented if all places with above-median access to community radio had had their own. While the effects appear seven months after the campaign announcement, or four months after its mass scale-up, these are interpreted as intention-to-treat effects. Effects materialize within two months of a station's effective adoption of the campaign, and they grow with exposure (Section \ref{sec:events_dcdh}).





Perhaps surprisingly, the impact of community radio is not driven by the shared ethno-linguistic makeup of its listeners: locality, more than ethnicity, defines the relevant community (Section \ref{sec:emp_hyp}). The effects are also not driven by overall access to community radio, indicating that informational content alone is not driving impacts: content is fixed for anyone with a specific community radio signal. National radio has no effect, further validating the design.

The results are robust to a battery of checks (Section \ref{sec:robust}). Local-radio access is uncorrelated with demographic, economic, infrastructure, and trust covariates conditional on overall community-radio coverage. I address selection on unobservables by replicating the results excluding the location of the radio transmitter, or examining the intensive margin using variation in local coverage conditional on having a local station. A political-accountability channel \citep{besley2002political, eisensee2007news} is ruled out by controlling for the presence of Ebola Treatment Units and laboratories, and showing these are not correlated with local-radio access, and are not deployed in the months when the main impacts are observed. Results hold under lagged-dependent-variable models, alternative case definitions, alternative measures of social resistance and treatment uptake, and more parsimonious specifications. \\



\begin{figure}[h]
	\vspace*{-.4cm}
	\setstretch{1}\selectfont
	\caption{Ebola infections by access to local community radio \label{fig:did_main} }
	\vspace{-0.5cm}
	\center\includegraphics[width=.6\columnwidth, trim=4 10 4 1,clip]{\figures/03_rsmooth/rsmooth_RR_dist_\radio_lebolaM_allnw.pdf}
	\begin{threeparttable}
		\begin{tablenotes}
			\vspace{-0.5cm}
			\item[] \footnotesize \textit{Notes}: Average log-Ebola in locations (sub-prefectures) with above- and below-mean access to a local community radio station, conditioning on locations with above-median access to some community radio station. The log is taken over the number of infections+1 (alternative definitions in Figure C2). Vertical lines indicate when the public health campaign was launched on community radio (gray dashed line) and the scale-up of the campaign as significant international aid arrived, helping community radio boost capacity, increase the number of hours aired, and improve the quality of the sensitization campaign (red solid line). Because the outcome is in logs, differences between the two lines over time correspond to differences in the growth of infections.
		\end{tablenotes}
		\vspace*{-.1cm}
	\end{threeparttable}
\end{figure}

To understand the mechanisms through which media affects the spread of Ebola, I develop a conceptual framework of common knowledge under multiple public signals (Section~\ref{sec:concept}), which organizes two candidate channels through which locality could matter: local radio reaches the very community that observes and sanctions behavior (a channel of \textit{norm proximity}), or it may provide more relevant or more trustworthy content (\textit{informational precision or trust}). I then draw on several pieces of evidence to distinguish empirically between them (Section~\ref{sec:mech}). First, I study changes in two measures of social health behavior, using the same difference-in-differences design above: a measure of ``social resistance'' (from the French \textit{réticence}) |attacks against health workers, opposition to public health measures, and refusal of safe burials| collected by the Guinean Ministry of Health, and the share of Ebola deaths occurring in the community rather than in a treatment unit, a proxy for treatment uptake recommended by the WHO, who provided this data. This evaluates the hypothesis that local media affects socially sanctioned behavior. Second, I use detailed data from the International Federation of Red Cross and Red Crescent Societies (IFRC) on their staggered rollout of local mobile-radio shows and door-to-door social-mobilization campaigns across Guinea, and refusals of safe burials conducted by them. Exploiting the timing of these interventions in a dynamic difference-in-differences design \citep{de2024difference}, I estimate their impact on refused burials at the prefecture-month level. This tests whether local media helps coordinate a change in social community norms, more than private information, proxied here through door-to-door campaigns. Third, I exploit cross-sectional variation in access to radio combined with data from a representative survey of 2,466 individuals collected by the Guinean National Institute of Statistics at the end of the outbreak, examining how pre-existing access to different media outlets affects second-order beliefs formed ex-post during the Ebola outbreak, on one hand, and private or non-socially-sanctioned health behavior, on the other.


The mechanism evidence points to local media facilitating a coordinated shift in socially-sanctioned behaviors by reaching the very community that observes and sanctions one's behavior |i.e.\ \emph{norm proximity}, as opposed to \emph{informational precision or trust}. Local radio significantly raises adherence to socially-sanctioned health behaviors, reducing social resistance and shifting Ebola deaths away from the community toward treatment units (Section \ref{sec:mech_social}). Local mobile radio shows aired by the Red Cross are followed by a reduction in refused safe and dignified burials, while door-to-door campaigns are not, indicating that the common knowledge aspect of media is key to achieving a coordinated shift in locally-sanctioned behavior (Section \ref{sec:mech_ifrc}). Survey evidence on second-order beliefs corroborates the coordination mechanism. People who learned about Ebola through media hold stronger beliefs that others in their locality will seek treatment when sick, and the magnitude of this effect grows with the proportion of people in one's locality who heard about Ebola through media |indicating media acts as a coordination device. Media has no detectable effect on private behaviors such as washing one's hands or clothes with chlorine, beyond other information sources (Section \ref{sec:mech_beliefs}). Because informational precision or trust should affect private behavior as well, this pattern rules them out as primary explanations. Heterogeneity by pre-existing prefecture-level trust in media (Afrobarometer 2013) provides further, if indirect, evidence against trust as the source of the locality advantage: if local radio mattered because its own community trusts it more than other stations, its impact should vary with trust in media, and it does not (Section \ref{sec:mech_trust}).

The findings advance our understanding of how community norms change, with implications not only for pandemic containment or vaccine uptake, but also for changes in gender norms, or norms around corruption and harassment. Even in highly ethnically diverse contexts, where ethno-linguistic identity is often assumed to define the relevant group, a relevant community can be the local one that observes and sanctions behavior. Local media can drive collective shifts in norms by reaching that community. 


\section*{Contribution to Literature}



The main contribution of this paper is to identify the impacts and channels through which local media shapes community norms. I build on a literature on the economics of media \citep[reviewed by][]{stromberg2015media, dellavigna2015economic}.\footnote{This literature studies media impacts on a variety of socio-economic behaviors, including civic engagement, political attitudes, gender norms, violence, and demonetization \citep[e.g.][]{stromberg2004radio, dellavigna2007fox, ferraz2008exposing, jensen2009power, olken2009television, paluck2009deference, enikolopov2011media, card2011family, durante2012partisan, ferrara2012soap, dellavigna2014cross, yanagizawa2014propaganda, adena2015radio, durante2018attack, blouin2019erasing, green2020countering, armand2020reach, banerjee2024less}.} Previous work found a role for non-national or private media \citep{grosfeld2024independent, pinna2022cable, buhler2026couch}, for community radio \citep{keefer2011mass, keefer2014mass, glennerster2026media}, and for media with shared language or ethnicity \citep{oberholzer2009media, wang2020black}. I contribute to this work in three ways: by holding the information source and contents constant, I identify the effects of media being local; by providing first evidence differentiating between locality and ethno-linguistic groupings, showing that locality matters beyond ethnicity; and by identifying norm proximity as a key channel through which local media helps shift social behavior, formalized in a conceptual framework.\footnote{The model of common knowledge under multiple public signals builds on a global games literature \citep{morris2001global, morris2002social, cornand2008optimal, kuznetsova2016value, shadmehr2020coordination}.}


I add to the literature on social norms and cultural change \citep{bisin2011economics, gershman2017long, bursztyn2018misperceived, bursztyn2020extreme}, to research on the role of information in facilitating these shifts \citep{karing2024social, gulzar2025good}, and to work on the mechanisms behind harmful community norms \citep{de2019does, corno2020age, montero2022religious, gulesci2025stepping}. I contribute to this evidence by identifying the role played by local media in shifting community norms. The rapid change in cultural practices facilitated via local media is consistent with recent evidence on the role of norms in explaining attitudes towards social distancing during the Covid-19 pandemic \citep[e.g.][]{martinez2021let, allen2024correcting}. The findings also underscore the importance of cultural diversity beyond ethnicity in shaping culture \citep{alesina2002trusts, desmet2017culture}.

This work adds to our understanding of health behavior in developing countries. Distinguishing between social and private health behavior reconciles mixed evidence on the impacts of information on health behavior, ranging from no effect \citep{kremer2007illusion, banerjee2010improving, duflo2015education, dupas2024negligible}, to an effect depending on the setting \citep{dupas2011teenagers, galiani2016promoting, banerjee2019entertaining, banerjee2024messages, breza2021effects, yang2023knowledge}. It also adds to the literature on institutional factors and the political economy of epidemics \citep{adda2016economic, richards2015social, christensen2021healthcare, van2020traditional, brodeur2021literature, lowes2021legacy, maffioli2021political, martinez2022vaccines, campante2024virus, gonzalez2025epidemics}. \\



Section \ref{sec:data} describes the background and data on the Ebola outbreak, Guinea's media landscape and public health campaigns. Section \ref{sec:concept} describes the conceptual framework. The main empirical set-up and results are presented in Section \ref{sec:empirical}. Empirical evidence distinguishing between different mechanisms is provided in Section \ref{sec:mech}. Section \ref{sec:conclusion} concludes. 

\section{Background and Data}\label{sec:data}

I combine a rich amount of data on the Ebola outbreak, public health response, and media landscape in Guinea from own-conducted surveys and secondary sources obtained in Guinea, summarized here, with additional details in Online Appendix (OA) Section D on data construction, Section E on background on the Ebola outbreak, and Section F on radio programming. 

\subsection{Epidemic disease, socially-sanctioned behavior, and other beliefs and behavior} \label{sec:data_out}

The 2014 West African Ebola epidemic is the largest Ebola virus disease (EVD) outbreak to date, causing over 28,600 infections and over 11,300 deaths. It hit West Africa for the first time in the known history of the disease at the end of December 2013 in Guéckédou in Nzérékoré, Guinea, at the borders of Liberia and Sierra Leone. The first infection was caused by the rare circumstance of an animal-to-human contagion, likely resulting from the contact of an 18-month-old boy with a bat. Importantly, all subsequent cases spread exclusively through human-to-human contact. The West African epidemic came to an end mid-2015, except in Guinea, which saw a lower Ebola burden compared to neighboring Liberia and Sierra Leone, but a longer outbreak, with a significant number of cases throughout 2015, ending in April 2016. Ending it required treating symptomatic individuals, isolating infected people, tracing their contacts, ensuring safe burials and changing population behaviors towards protective habits.\footnote{OA Section E gives additional background on the Ebola outbreak.}

\paragraph{Ebola infections.} 
The main outcome is the number of Ebola infections per month per sub-prefecture, obtained from patient records from the World Health Organization (WHO) in Guinea, and measured from January 2014 to June 2016. There were a total of 3,814 (2,544) Ebola infections (deaths) in Guinea (see Table C1, Figure C1).\footnote{The baseline measure includes probable and confirmed cases. This is consistent with the reporting practice of the WHO, which publishes weekly data of confirmed and probable cases. These have been evaluated by a clinician or have a link to a confirmed case, where confirmed cases in addition have a positive laboratory result (see Table D1 for definitions). I also validate results with confirmed cases only, and with Ebola deaths.} Ebola is short-lived, with people typically dying within four weeks since contagion, and a fatality-rate that widely varies from 25-90\%, depending on treatment-uptake.\footnote{Treatment-uptake and improved public health systems lowered Ebola death rates \citep{garske2017heterogeneities}.} Contagion occurs only with advanced symptoms, through the exchange of bodily fluids. This implies that health behavior can significantly affect death rates and reduce infections. Ebola survivors suffer from persistent medical conditions after recovery, including loss of sight, mental illness, and increased mortality risk \citep{james2020}.  


\paragraph{Socially-sanctioned behavior.} Two main measures of health behavior are observed at the sub-prefecture level over time; a third is available at the prefecture level for a shorter period.

\subparagraph{(1) Social resistance.} The Guinean Ministry of Health tracked instances of ``social resistance'' (from the French, \textit{réticence}), such as attacks against public health authorities, refusing treatment, or opposition to safe burials for family members. These were opposed as they countered cultural practices and prevalent beliefs about the disease.\footnote{Reported attacks against Red Cross volunteers, who were in charge of burials, averaged ten per month in the second half of 2014 (\href{http://www.ifrc.org/en/news-and-media/press-releases/africa/guinea/red-cross-red-crescent-denounces-continued-violence-against-volunteers-working-to-stop-the-spread-of-ebola/}{IFRC website}). Interviews with social workers in Guinea reveal that ETUs were a source of conspiracy theories. These were established ad-hoc, as Ebola patients need to be isolated to avoid contagion. People had to refrain from taking care of sick family members and saw them dying in ETUs at high rates.} Data on such instances was published in daily or weekly situation reports, which I digitized using data-scraping algorithms (see Figure C3). As baseline measure I compute the number of weeks in a month with at least one such resistance behavior for a given sub-prefecture (taking values from 0-5). 


\subparagraph{(2) Treatment uptake.} I compute the number of people who die from Ebola in their own community divided by the total number of Ebola deaths, which include those dying in an Ebola treatment unit (ETU), established ad-hoc to isolate and treat Ebola patients. This is a proxy for low treatment-uptake or opposition to public health measures proposed by the WHO in Guinea, who provided this data. Individuals who oppose treatment refrain from going to an ETU; if they die, they are more likely to do so within the community than in an ETU.

\subparagraph{(3) Refusal of ``safe and dignified burials'' (SDBs).} 	Burial practices, a central cultural and religious practice, were the most significant factor contributing to the spread of Ebola in Guinea.\footnote{Nearly $60\%$ of all Ebola cases in Guinea can be linked to traditional burial practices (\href{https://www.who.int/news-room/spotlight/one-year-into-the-ebola-epidemic/factors-that-contributed-to-undetected-spread-of-the-ebola-virus-and-impeded-rapid-containment}{WHO January 2015}).} Changing these practices constituted an important source of distress and was often met with opposition. In Guinea, the Red Cross carried out safe and dignified burials (SDBs), known as such as they were gradually adapted to the local traditional and religious customs. I measure communities’ refusal to these burials. The data was obtained from the International Federation of Red Cross and Red Crescent Societies (IFRC) in Guinea and includes information on SDBs, refusals of burials, and other activities conducted by the organization, including mobile radio shows and social mobilization campaigns. The data is available by prefecture per week from April to December 2015, which I aggregate at monthly level.	


\paragraph{Second-order beliefs and private behavior.} A cross-sectional representative Ebola survey with 2,466 individuals conducted by the Guinean National Institute of Statistics (INS) in September 2015 provides information on other health behaviors and beliefs about Ebola (Section \ref{sec:mech_beliefs}). 


\paragraph{Containment policies, other diseases.} The presence of ETUs, laboratories, and community care centers (CCCs), are sourced from WHO, UNOCHA/UNMEER. Data on the prevalence of other infectious diseases by prefecture for 2013-2014 are from the Guinean Ministry of Health.


\subsection{Media landscape and public health campaigns \label{sec:data_treat}}

Information on public health campaigns on different media outlets was collected during in-person and phone interviews, as well as through surveys with radio stations across Guinea. The main public health campaign aimed at curbing Ebola was aired on community radio. Other radio stations, which include the national, private, and international radio, provided some public health information, but this was either late and short-lived or limited to descriptive information about Ebola. This is summarized in Table A6, with further details in OA Sections D.2 and F. 

\paragraph{Public health campaign.}  

The main public health campaign was a coordinated effort from community radio stations, yet delivered by local hosts in local languages across the country. It was aimed at convincing the population to avoid taking care of sick family members, seek health treatment instead, and change traditional funeral practices, the main causes of the disease spreading. It was officially launched at the end of June 2014, but most community radio stations started the program after September 2014, when the non-governmental organization \textit{Fondation Hirondelle} provided technical assistance and created content in multiple local languages, which radio stations uploaded into their programs. Over time, the focus shifted from a descriptive account of the disease, at times contradictory as medical knowledge evolved, to focusing on high-risk habits, such as traditional burial practices and low treatment-uptake.\footnote{I have information on community radios' adoption date of this second phase of a fully-fledged sensitization campaign for a subset of community radio stations (used in Section \ref{sec:events_dcdh}).} In this phase community radio stations invited local Ebola survivors, traditional or religious leaders, and directly addressed prevalent rumors (see Table A6, details in OA Sections D.2 and F).

\paragraph{Community radio, locality, and ethno-linguistic belonging.} 22 out of the 34 prefectures had their own community radio prior to the outbreak. Together, they form the \textit{Radios Rurales de Guinée (RRG)}, and they have a coordinating body in the capital, Conakry. Each community radio was established in the past to serve a given prefecture, including in terms of contents and languages spoken. I define community radio as \textit{local} to all sub-prefectures belonging to the prefecture where a community radio is based, as proposed by the director of RRG. Since the radio signal can reach beyond administrative borders, some sub-prefectures have access to a community radio station that is not local to them. In addition, to study the impacts of ethno-linguistic belonging, as opposed to the locality, I define the majority ethno-linguistic group of a community radio by the ethnic majority of the prefecture it serves, which also coincides with the main languages reported by the radio stations I surveyed. Finally, when policy makers saw the apparent success in curbing Ebola, \textit{RRG} decided to open 10 new community radio stations in the remaining prefectures except for two.\footnote{Except Coyah, the sub-urban outskirts of Conakry, and Lola, a small prefecture east of Nzérékoré.} Since these were launched at the end of the baseline study period, I only consider them as treated in a complementary analysis using the actual date of adoption of the public health campaign for the subset of community radio stations for which I have their actual launch date (Section \ref{sec:events_dcdh}).\footnote{2 prefectures opened a new community radio station in April 2015, 1 in May 2015, the remaining ones after August 2015, and I have the month of launch for 7 out of 10 prefectures (see OA Sections D.2 and F). Only 3 of these prefectures had 1-8 active Ebola cases when these were launched.\label{foot:newradio}}

\paragraph{Radio signal reception.} To measure radio access, I use data on 155 radio transmitters from the National Telecommunications Agency of Guinea (ARPT), combined with information on geographic terrain features. I calculate the radio signal reception using a radio propagation tool employed by engineers, based on the Irregular Terrain Model / Longley-Rice, standard in the economics literature since \citet{olken2009television}. Guinea's ruggedness is among the top half of African countries \citep{nunnpuga2012}, offering quasi-random variation in radio access.\footnote{Guinea's elevation spans from 0 to 1752 meters (5,748 feet) above sea level. Ruggedness is not as high as in Rwanda, but similar to Indonesia, and twice that of Congo, where this method was used to study impacts of media \citep{olken2009television, yanagizawa2014propaganda, armand2020reach}.} I group radio transmitters by the type of radio and obtain the share of a given sub-prefecture with access to a given radio station with the signal strength required for normal receivers, and a lower threshold for robustness ($43$ and $33 \; dB \mu V/m$, respectively, shown in OA Figures C4-C5).\footnote{These thresholds are set by the Nautel Radio Coverage Tool online software used to compute radio signal reception. This radio coverage is then parsed in Python to compute the overlap of different radio signals, and ultimately the share of a sub-prefecture with access to a given type of radio station.}

Most sub-prefectures have at least half of their territory covered with access to some radio (Table A2). The most common type of radio is community radio, followed by national radio. The mean (median) share of a sub-prefecture with \textit{its own} community radio is $27\%$ ($10\%$).

\paragraph{Radio listening patterns.} Data on whether an individual has learned about Ebola on the media, and whether they own a radio device, are obtained from the Ebola INS survey (2015) described at the end of Section \ref{sec:data_out}. I also use information on respondents' preferred radio station, available only after the end of the outbreak, from Afrobarometer round 7 (2017).

\subsection{Geography, ethno-linguistic diversity, and demographic data} 

Guinea has a total of 34 prefectures, split into 340 sub-prefectures with an average population of 33,000 each, whose geographical features I obtain from geoBoundaries. Administrative borders relate to ethno-linguistic groupings, but still harbor important diversity (see Section \ref{sec:H1b_ethn}). Demographic data are from the 2014 Guinean census, and from Afrobarometer round 5 (2013) (see Tables C1 and A5, Figure C7).


\section{Conceptual Framework}\label{sec:concept} 

How does public information, as opposed to private information, impact behavior, and why may the locality of media matter for affecting different types of behavior? I first present a framework, formalized in OA Section~B, that clarifies (i) what distinguishes \textit{public} from private information, and why this matters for affecting \textit{different types of behaviors}; and (ii) why the \emph{locality} of a media outlet may matter, through three distinct channels with distinguishable empirical predictions (Section \ref{sec:concept_model}). I then discuss how this behavior affects the spread of disease based on epidemiological theory (Section \ref{sec:epid_model}). The framework directly motivates the working hypotheses tested in the paper (Section \ref{sec:hypotheses}).

\subsection{How radio impacts different behaviors \label{sec:concept_model}}

\paragraph{Public information and socially-sanctioned behavior.} The defining feature of radio, relative to private information channels (such as door-to-door visits, advice from a doctor), is that it is \emph{common knowledge}: each listener knows that others heard the same message. Common knowledge helps achieve coordinated shifts in behaviors that are observed and sanctioned by the community \citep{morris2001global, ferrara2012soap, dellavigna2014cross, yanagizawa2014propaganda, cantoni2017protests, armand2020reach, green2020countering}. During the Ebola epidemic, the most consequential cultural norms ---traditional burial practices and norms around caring for the sick--- were precisely of this kind, and required a rapid coordinated shift. For private behaviors that are unobserved or not reprimanded, such as washing hands or clothes with chlorine, the common-knowledge property adds little; there may even be free-riding once others are known to be taking precautions \citep{lau2019social, allen2024correcting}. This yields a first prediction tested in the paper: \emph{Public information shifts socially-sanctioned behaviors more than private information does. For behaviors that are private or not reprimanded, the public–private distinction matters little |and may reverse with free-riding.}

\paragraph{Why locality matters: three channels.} Whether the \emph{locality} of radio matters speaks to who is the relevant ``other'' whose behavior and beliefs influence one's own. To organize this, I extend \citet{morris2002social} and \citet{cornand2008optimal} to a setting with multiple public signals (OA Section~B). Locality enters through three distinct channels:

\begin{enumerate}[leftmargin=*, itemsep=2pt]	
	\item \textbf{Contagion proximity:} listeners closer to the epidemic have a stronger private incentive to respond to \emph{any} radio signal. This channel does not differentiate local from non-local radio, since this incentive is common across radio signals for a given listener.
	\item \textbf{Norm proximity:} a station based in one's own community broadcasts to the same people who observe and sanction one's behavior, so the cost of deviating from the new norm is high. A non-local station, even with the same content, reaches a more diffuse audience that cannot enforce local norms, and therefore lacks the same coordinating effect.
	\item \textbf{Informational precision and trust:} local radio may provide more relevant, specific, and linguistically appropriate content about the local epidemic; listeners may also trust their local station more. These two factors jointly determine how much listeners will update their beliefs, and are observationally equivalent when testing behavior-type patterns alone.\footnote{I examine the trust channel separately, using heterogeneity by pre-existing trust in radio and stated preferences for community radio over other stations (Section~\ref{sec:concept_dist}, Appendix~B.2.2).}
\end{enumerate}

\paragraph{Who is the relevant ``other'': locality vs.\ ethnicity.} A natural alternative to geographic locality is ethno-linguistic identity: the literature documents pervasive ethnic divides in Africa \citep{fearon2003ethnicity, alesina2005ethnic, michalopoulos2016long} and a role for media along ethno-linguistic lines in the US \citep{oberholzer2009media, wang2020black}. It is possible that the community that sanctions behavior is that of ones' own ethnic group. Yet cultural norms also form beyond ethnicity \citep{fearon1996explaining, alesina2002trusts, desmet2017culture}. Since behaviors can only be sanctioned by people who observe them, and observation of health behaviors requires physical proximity, the model suggests the relevant ``other'' is geographically local \citep[e.g. consistent with conformity in mask wearing during Covid-19 as a function of close contacts, ][]{woodcock2021role}. I measure locality at the prefecture level, the administrative unit at which community radio stations are based prior to the epidemic (see Section~\ref{sec:data_treat}).\footnote{Prefecture origins predate the colonial period and partly map onto ethno-linguistic groupings \citep{goerg2011couper, mcgovern2012unmasking, gadenne2014decentralization}. This community-based logic is also consistent with sociological evidence on community-based campaigns \citep{katz1966personal, richards2016ebola} and with epidemiological models of overlapping social networks \citep{funk2009spread}.} Ultimately, what is the relevant community is an empirical question.

\subsection{How radio impacts epidemic spread \label{sec:epid_model}}

	The equilibrium degree of protective behavior in turn affects the transmission rate of Ebola. Following the epidemiological literature, new infections in sub-prefecture $s$ at month $t$ follow exponential growth with transmission rate $\beta_{s,t}$ on a fraction $\rho \in [0,1]$ of past cases \citep{lekone2006statistical, fang2016transmission}, where the public health campaign affects $\beta_{s,t}$ through equilibrium behavior \citep{funk2009spread, kiss2010impact}: $Ebola_{s,t} = Ebola_{s,t-1}^\rho \; e^{\beta_{s,t}}$. Taking logs and approximating $\beta_{s,t}$ as a linear combination of the baseline transmission rate, the public health campaign, and covariates, gives the equation that guides the main empirical strategy:\footnote{Transmission rate $\beta_{s,t}$ is a function of baseline transmission rate $\beta_0$, equilibrium strategies $b(P_{s,t})$ and baseline covariates $X_{s,t}$, where the equilibrium strategies are impacted by access to the campaign $P_{s,t}$.}
	\begin{align}
		logEbola_{s,t} =\rho \; logEbola_{s,t-1} + \beta_0 + \beta_1 \; P_{s,t} + \beta_2 \; X_{s,t}	+ u_{s,t} \label{eq:theory2}
	\end{align}
where $P_{s,t}$ is access to the public health campaign (the share of sub-prefecture $s$ covered by a given radio station from the campaign's launch onwards, zero before), $X_{s,t}$ are baseline covariates, and $u_{s,t}$ a random component (e.g.\ contagion with distant communities).

\subsection{Working hypotheses on impacts of different radio stations \label{sec:hypotheses}}

The framework above poses two sets of questions, taken to the data sequentially. First, what are the impacts of radio, of local radio, and what is the relevant ``local'' |geography, ethno-linguistic identity, or a combination? I lay out here the working hypotheses, evaluated in the main empirical section (Section~\ref{sec:empirical}). Second, once established what is ``locality'', and that local radio impacts the epidemic disease, I ask why locality matters |through norm proximity, or through informational precision or trust? These generate distinguishable predictions across behavior types and across heterogeneity by trust, evaluated in Section~\ref{sec:mech}.

\begin{figure}[ht]
	\setstretch{1}\selectfont
	\center
	\centering
	\caption{Stylized map illustrating reach of community radio in prefecture A, and working hypotheses}
	\label{fig:fakemap}
		\includegraphics[width=\columnwidth, trim= 0 10 0 0, clip]{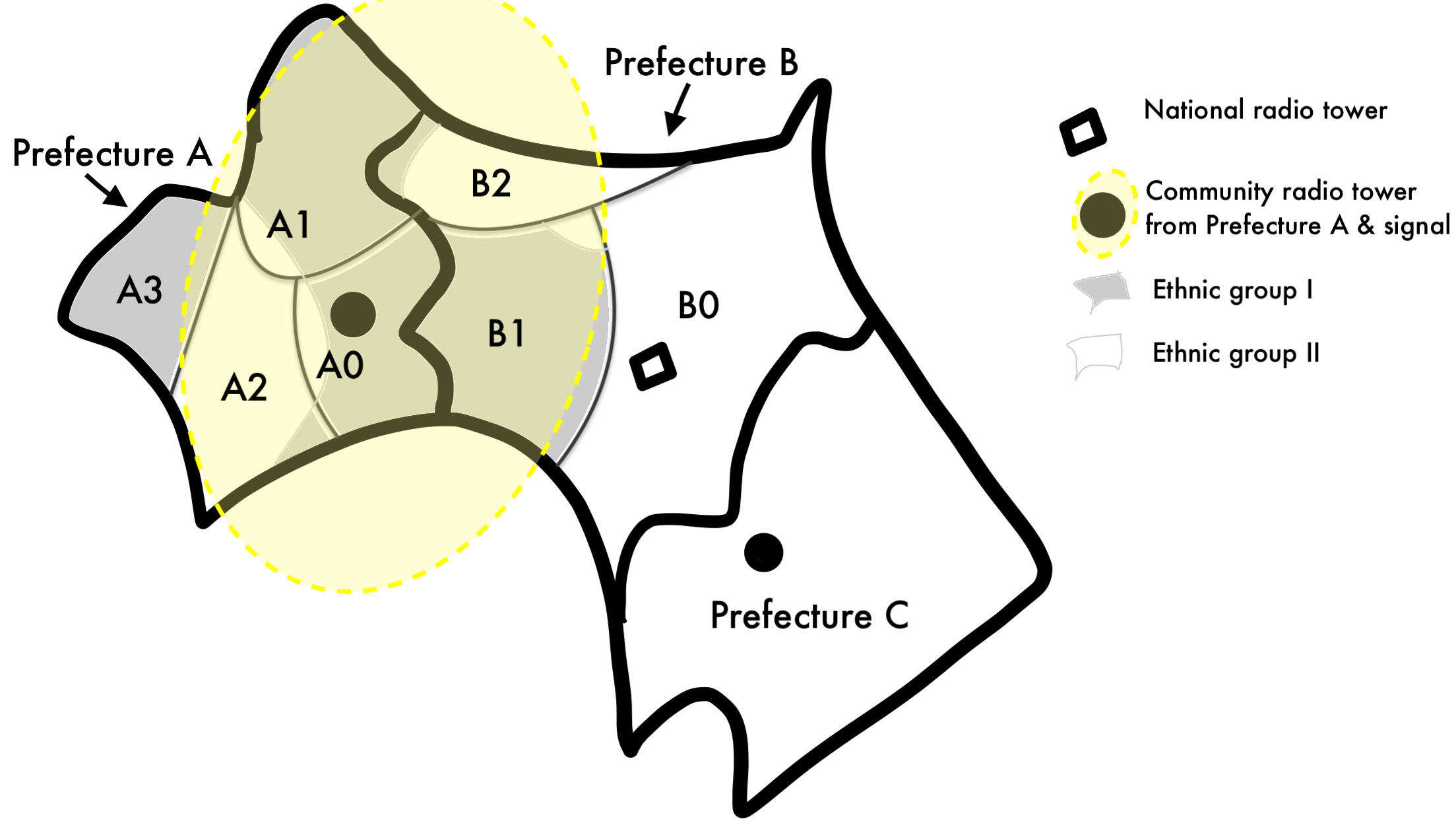} 
		\small
		\begin{threeparttable}
			\vspace{-.5cm}
			\begin{tablenotes}\footnotesize
				\item[] \footnotesize \textit{Notes}: The Figure presents a stylized map of Guinea, illustrating different hypotheses described in Section \ref{sec:hyp}. Black circles (diamond) symbolize the location of the radio transmitter of a community (national) radio station; bold (thin) black lines are prefecture (sub-prefecture) borders; gray and white shapes inside sub-prefectures symbolize different ethnic groups; the large oval shape with a dashed line symbolizes the radio signal corresponding to the community radio placed in prefecture A (sub-prefecture A0).
			\end{tablenotes}
		\end{threeparttable}
\end{figure}

\subsubsection{Main hypotheses | evaluating impacts of radio and ``locality''}\label{sec:hyp}

 I lay out the corresponding hypotheses with the help of the stylized map in Figure~\ref{fig:fakemap}. Assume there are at least two prefectures, two ethnic groups, and at most one community radio station per prefecture. Focusing on the community radio in prefecture A (sub-prefecture A0), I ask which sub-prefectures would see an earlier drop in infections, for a given channel at work. This helps define the empirical test used in Section~\ref{sec:empirical}. I will compare sub-prefectures with greater radio access, to others, evaluate pre-trends, and examine whether they see an earlier drop in infections in the months after the public health campaign launch.

\begin{itemize}[leftmargin=*, itemindent=0pt, itemsep=0pt]\setlength{\itemsep}{0 mm}
	\item[] \textbf{Hypothesis 1:} Information from \textit{any community radio} impacts people across the country, leading to lower infections over several months. \textit{Empirical test:} Sub-prefectures with greater access to any community radio (A0, A1, A2, B1, B2) see an earlier drop in infections. 
	
	\item[] \textbf{Hypothesis 1a (main hypothesis):} Information from \textit{community radio} impacts people in the \textit{locality} (prefecture) it serves. \textit{Empirical test:} Sub-prefectures covered by their own community radio (A0, A1, A2) see an earlier drop than others, including those covered by community radio from other prefectures (B1, B2).

	\item[] \textbf{Hypothesis 1b:} Information from community radio impacts people from any locality but with \textit{shared ethnic identity} as the majority in the prefecture the community radio serves. \textit{Empirical test:} Sub-prefectures covered by community radio of any prefecture with the same ethnic majority (A0, A1, B1) see an earlier drop than others, including those covered by community radio from prefectures with a different ethnic majority (A2, B2).

	\item[] \textbf{Hypothesis 1c:} Information from community radio impacts people in the \textit{locality} it serves, especially for those who \textit{share the ethnic identity} of the majority in that prefecture. \textit{Empirical test:} Sub-prefectures covered by their own community radio (A0, A1, A2) see an earlier drop than others, and this drop is driven by sub-prefectures that share the ethnic majority of the radio's prefecture (A0, A1, as opposed to A2).
	
	\item[] \textbf{Hypothesis 2:} Information from \textit{any radio} (e.g.\ national radio) impacts people across the country, leading to lower infections over several months. \textit{Empirical test:} Sub-prefectures with greater access to national radio see an earlier drop in infections.
\end{itemize}

\paragraph{Mapping hypotheses to model channels.} \textit{Hypothesis~1a} is the main hypothesis: the locality channel of the model. The remaining hypotheses test alternatives. \textit{Hypothesis~1} (any community radio) speaks to whether the contents of the public health campaign designed by community radio are sufficient to influence behavior, regardless of locality. \textit{Hypothesis~1b} tests whether the relevant ``other'' is the ethno-linguistic group rather than the geographic community. \textit{Hypothesis~1c} evaluates the role of ethnicity within a locality. \textit{Hypothesis~2} serves as a falsification test for access to ``any radio'' driving impacts. I evaluate these in Section \ref{sec:empirical}.

\subsubsection{Disentangling mechanisms | why ``locality'' matters}\label{sec:concept_dist}

Different mechanisms through which local radio may impact behavior |norm proximity, or informational precision and trust| generate distinguishable predictions (see OA Section B.2.3). If \textit{informational precision or trust} is the primary channel, local radio should outperform non-local radio for both private and socially-sanctioned behaviors, since both benefit from relevant and trustworthy information. If \textit{norm proximity} is the primary channel, local radio should have no advantage for private behaviors, which do not counter social norms, and a positive advantage for socially-sanctioned behaviors, which require a coordinated shift in behavior. A null effect of local radio on private behavior therefore rules out informational precision and trust as the primary explanations, pointing to norm proximity as the key channel. To \textit{distinguish between trust and informational precision} empirically, I evaluate whether there are heterogeneous effects of local radio by pre-existing levels of trust in radio, and stated preferences for community radio over other stations. I examine this in Section~\ref{sec:mech}.

\begin{figure}[H]
	\setstretch{1}\selectfont
	\center
	\centering
	\caption{Ebola burden and access to community radio \label{fig:eb_prepost}}
	\vspace{-0.2cm}
	\hspace{2cm}\includegraphics[width=.75\columnwidth, trim=0 10 0 10, clip]{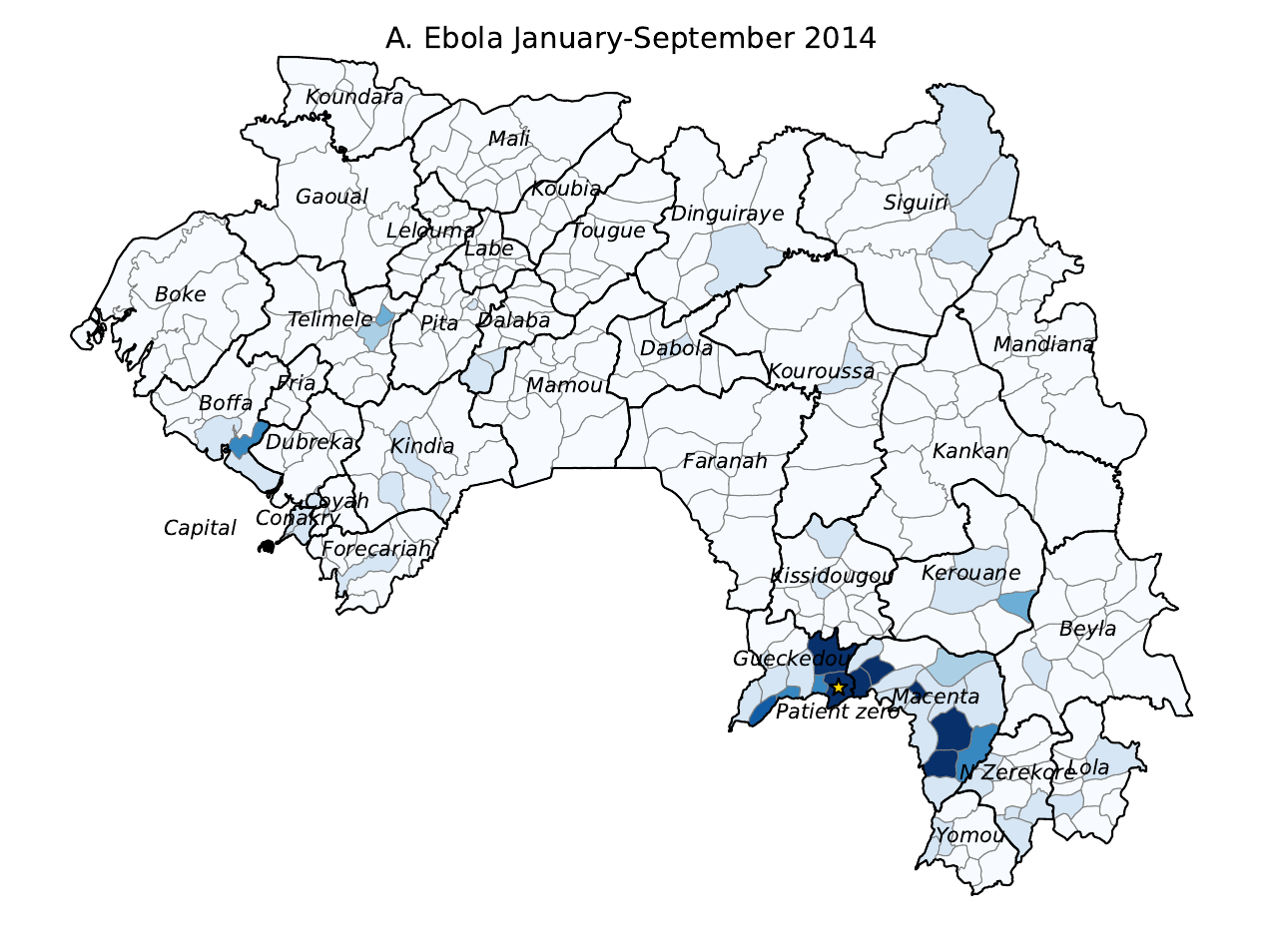} \\
	\vspace{-2.4cm}
	\hspace{-2cm}\includegraphics[width=.75\columnwidth, trim=0 10 0 10, clip]{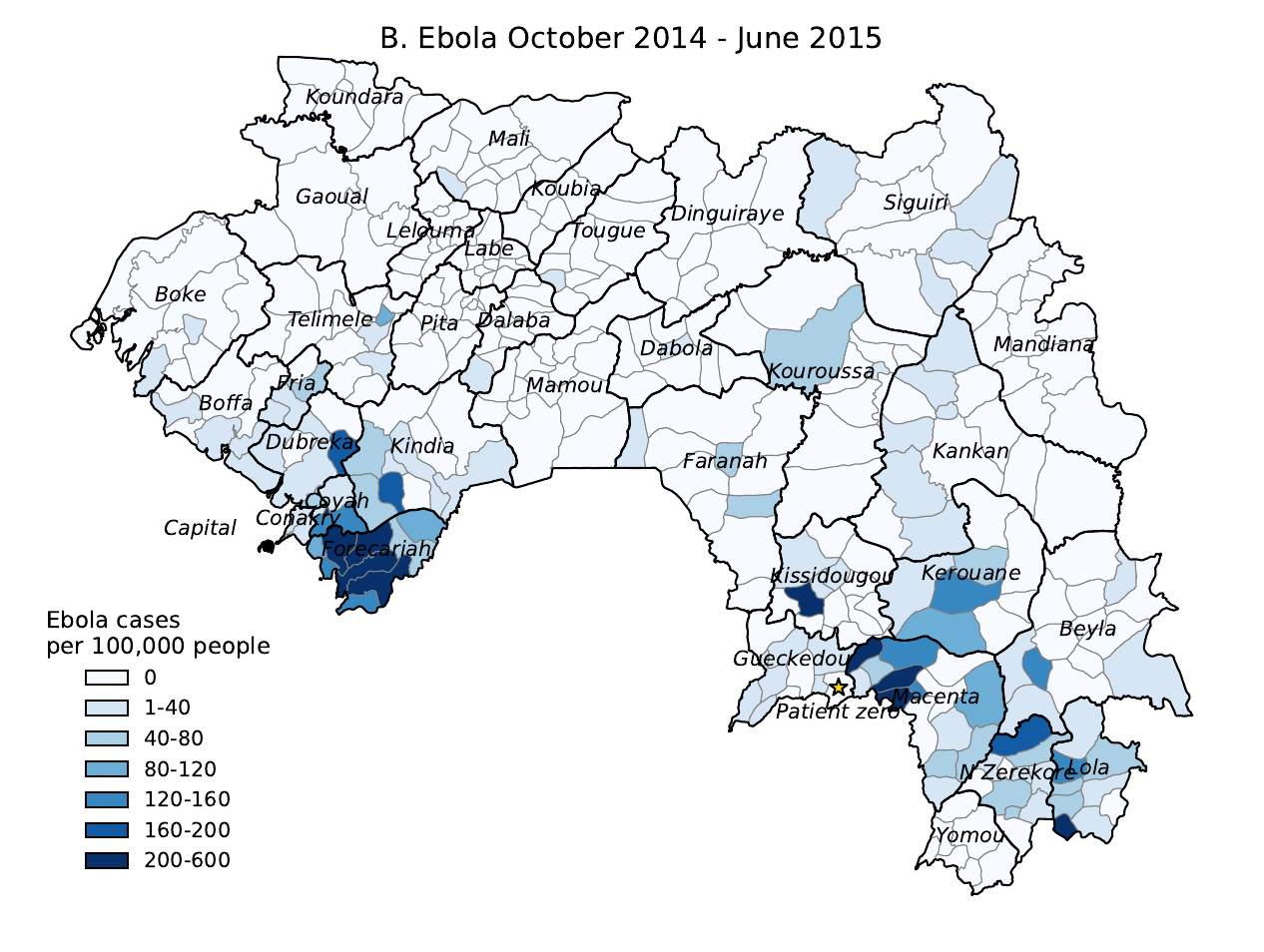} \\
	\vspace{-0.3cm}
	\includegraphics[width=.75\columnwidth, trim=0 10 0 10, clip]{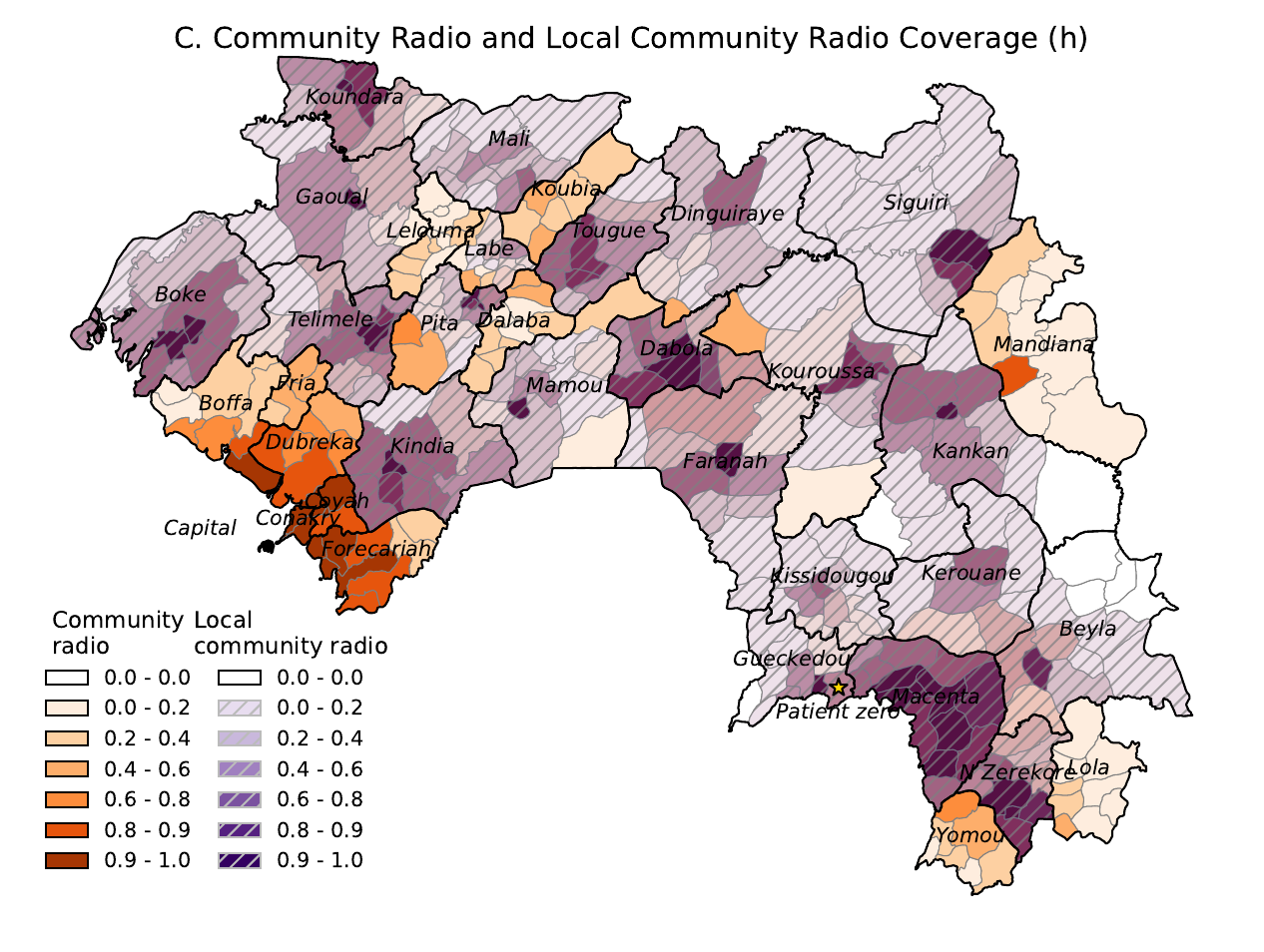}
	\begin{threeparttable}
		\vspace{-.7cm}
		\begin{tablenotes}
			\setstretch{1}\selectfont
			\item[] \footnotesize \textit{Notes}: Cumulative number of Ebola infections per 100,000 people in a sub-prefecture before (panel A) and after (panel B) the public health campaign on community radios. Share of a sub-prefecture with access to a community or to a local community radio station, i.e. one based in their own prefecture (panel C). For health facilities and population see Figure C6.
		\end{tablenotes}
	\end{threeparttable}
\end{figure}

\section{Empirical set-up \label{sec:empirical}}

\subsection{Motivating evidence}

Figure \ref{fig:eb_prepost} motivates the main hypothesis and empirical strategy. The upper panels show cumulative Ebola cases before and after September 2014, when most community radio stations launched the public health campaign. The lower panel shows access to local (solid purple) and non-local (dashed) community radio. Two pairs of prefectures illustrate the pattern. In the south-west, Forécariah and Kindia were both unaffected before October~2014 and had similar overall community-radio access, but only Kindia had its \emph{own} local station |and saw a much lower subsequent epidemic burden, despite Forécariah being more populated and having no worse access to health facilities (Figure~C6). In the south-east, Nzérékoré had a local community radio while neighboring Lola did not, and sub-prefectures in Lola show a higher per-capita burden despite both being close to the epicenter, or first infection. The empirical design formalizes these contrasts: it compares the differential change in epidemic cases across places that all have access to community radio but differ in whether the station is their own \emph{local} one. 

\paragraph{Radio coverage and listening.} The design relies on radio coverage being a meaningful proxy for listening. Three pieces of evidence support this: it is positively and statistically correlated with individuals' stated preferred radio station being a community radio (Table A7, Panel A); with individuals having received information about Ebola from the media, owning a radio device, and the interaction of both (Panel B); and these correlations hold in a regression framework with controls (Table A12).\footnote{Suggestively, treatment effects are also larger where people report preferring community radio (Section \ref{sec:mech_trust}).} 


\subsection{Empirical strategy}

The main empirical strategy follows the logic of a difference-in-differences (DiD) strategy with continuous treatment. I study the evolution of the epidemic in places with varying access to a given radio station, several periods before and after the start of a major public health campaign. The strategy is summarized in equation (\ref{eq:flex_eb}):\begin{align} \label{eq:flex_eb} 
		y_{s,t} &= \sum_{\tau} \beta_\tau \; RadioShare^{k}_{s} \times \mathbbm{1}{_t(\tau)} 
		+ \textbf{X}_{s,t} \; \Gamma +\alpha_s +\lambda_t + u_{s,t}    
	\end{align}
	
	The baseline outcome $y_{s,t}$ is the log-number of Ebola-infected cases per 100,000 people in sub-prefecture $s$ at month $t$.\footnote{I use $log(Ebola+0.01)$; the coefficient $\beta$ can be interpreted as the $\%$ change in the number of Ebola cases as a result of a $1 \;pp.$ increase in access to radio; in the text I report effects of a $10 \;pp.$ increase as $10\beta$, a linear approximation whose exact counterpart, $100(e^{\beta/10}-1)$, is slightly smaller for the largest estimates (e.g. 24\% instead of 27\% for $\beta=-2.7$). Results are robust to alternative specifications, and to adding a dummy variable for each zero observation. I also use socially-sanctioned behavior as an outcome in Section \ref{sec:mech_social}.} The specification is motivated by epidemiological theory (equation \ref{eq:theory2}).\footnote{The baseline specification does not include a lagged dependent variable, as its inclusion with fixed effects is problematic in short panels \citep{nickell1981biases}. Since we are dealing with one single epidemic outbreak, assuming that the impact of past Ebola cases on the effect of interest will be largely covered by location-fixed effects is reasonable, since the initial spread of Ebola is largely driven by fixed characteristics, i.e., the distance to the first case and population density. Results are confirmed in a lagged dependent variable specification (see Section \ref{sec:robust}).} The model is identical to a standard difference-in-differences equation, with time dummies $\mathbbm{1}{_t(\tau)}$, for each month $\tau$, before and after the launch of the campaign. The omitted category is $\tau=-1$, one month before its launch in June 2014. 
	
	The main treatment variable is the interaction between the share of a sub-prefecture $s$ with access to a given radio station $RadioShare^{k}_{s} \in [0,1]$, and the time dummies $\mathbbm{1}{_t(\tau)}$, for each month $\tau>0$ after the start of the campaign. The cross-sectional variable is a measure of radio signal exposure, constructed using the geographic features of the terrain and the technical characteristics of the radio tower. $\alpha_s$, $\lambda_t$ are location and time fixed-effects capturing time-invariant differences across locations, and common changes over time. Standard errors are clustered at prefecture-level, to account for spatial and serial dependence, and bootstrapped.\footnote{200 replications; results with analytic clustered standard errors are similar.}

		I control for a set of time-varying, or time-invariant controls interacted with time dummies: First, the distance to the closest community radio transmitter to absorb unobservables that may drive the decision to set up a radio station in a given location. Second, the shortest distance to public health facilities (ETUs, laboratories and CCCs), to ensure that impacts are not driven by a change in their supply. Third, other factors that may facilitate the spread of disease: log-population, area in square-km, and distance to the epicenter, i.e. location of first infection.

	To test the main hypothesis, that impacts are driven by local community radio (\textit{Hypothesis 1a}), I set $k=Local$.\footnote{I also check that results hold when conditioning on overall radio access (Figure~A10).} As a robustness check I also control for a dummy variable $Radio^{Local}_s \in \{0,1\}$ which is equal to 1 if the radio station is located in its own sub-prefecture (or prefecture) and 0 otherwise, interacted with the time dummies. This ensures the estimated effects of local radio are not driven by unobserved characteristics that fundamentally differentiate sub-prefectures (or prefectures) that have a local radio station from those that do not.

		The coefficients of interest are $\{\beta_\tau\}_{\tau\geq 0}$. They measure the impact of the public health campaign $\tau$ months after its launch, in places with greater access to a given radio station $k$, on the growth rate of Ebola. The key identifying assumption is that in the absence of the public health campaign, Ebola cases would have grown at the same rate in places with different levels of access to a local radio. Since Ebola spreads for at least six months before the official launch of the campaign, the DiD specification allows me to evaluate whether infections evolve in parallel prior to the campaign, in places with varying radio access.\footnote{I also examine whether socio-economic characteristics and the burden of other infectious diseases at baseline are predicted by radio access. While this is not strictly necessary in a DiD framework, it helps substantiate the hypothesis that Ebola would have continued evolving in parallel absent the campaign.} I examine this by evaluating whether $\beta_\tau$ is constant for all $\tau<0$ before the official launch of the public health campaign in June 2014, or until its effective adoption after September 2014.

	\subsection{Evaluating competing hypotheses}\label{sec:emp_hyp}
	
	\subparagraph{Overall community radio access (Hypothesis 1)}
	I evaluate whether the contents of the public health campaign, aired on any community radio, are sufficient to influence behavior \emph{regardless of locality}. To test this hypothesis, I use as treatment variable the share of a sub-prefecture with access to any community radio station, i.e. setting $k=Community$ in equation (\ref{eq:flex_eb}). Results presented in Figure \ref{fig:events_hyp}, Panel A, show significant pre-trends prior to the start of the campaign, followed by a rise, rather than a reduction in Ebola cases. This is expected, since overall access to radio, including overall access to community radio, is positively correlated with population, and other baseline characteristics (Tables A8-A9). It also suggests that the content of the public health campaign alone is not sufficient to influence behavior.

	\subparagraph{National radio (Hypothesis 2)}
	I also test an alternative hypothesis, that national radio leads to an earlier drop in Ebola infections, using as treatment variable the share of a sub-prefecture with access to national radio, i.e. setting $k=National$, in equation (\ref{eq:flex_eb}). Again, there is no evidence for this (Figure \ref{fig:events_hyp}, Panel B). Given that national radio stations did not have a sensitization campaign, but rather provided descriptive information about the spread of the disease, this exercise also serves as a falsification test.\footnote{They emitted factual information from about April 2014 ($\tau=-2$), four months into the epidemic (Table A6).} If access to radio is correlated with other aspects influencing the spread of Ebola over time, such as state capacity which may become more important as the disease advances, we may worry that the observed effects are due to the former, and not to the radio campaign. It is reassuring to find that access to any radio, such as national radio, is not predictive of a drop in Ebola infections. 

	\subparagraph{Local community radio (Hypothesis 1a)}

	To test the impacts of the \textit{locality} of community radio, I use as treatment variable the share of a sub-prefecture with access to its own community radio, i.e. setting $k=Local$ in equation (\ref{eq:flex_eb}). For several months we observe no difference in the rate of Ebola infections in places with and without a local radio station, and this is followed by a drop in Ebola infections about four months after the wide adoption of the sensitization campaign (Figure \ref{fig:events_hyp}, Panel C). The parallel trends assumption is robust to different specifications, with or without controls, and controlling for overall community-radio access (Figure \ref{fig:events_local}). Conditional on having access to any community radio, places with and without a local community radio are similar along observable characteristics (Tables A8-A9). Sections \ref{sec:event_local_eb}-\ref{sec:robust} delve deeper into this hypothesis; with mechanisms discussed in Section \ref{sec:mech}. 
	
	\subparagraph{Community radio, based on ethno-linguistic belonging (Hypothesis 1b)}\label{sec:H1b_ethn}
	Since there are at least 24 distinct languages spoken in Guinea, |at the extreme| it could be that the importance of local radio is simply due to citizens' inability to understand the radio from a neighboring community.\footnote{Though unlikely, since 81\% of Guineans speak at least 2 languages (Table A5, Panel C); groups are somewhat clustered across the four main regions, so community radios often reach neighboring prefectures with shared language (Figure C7).} To evaluate the role of \textit{ethno-linguistic belonging}, I examine whether having access to a community radio sharing the same ethnic majority group or language is associated with a drop in Ebola infections. I set $k=Ethnic$ in equation (\ref{eq:flex_eb}), where the cross-sectional treatment variable is the share of sub-prefecture $s$ covered by community radio stations sharing the sub-prefecture's majority language. I find no evidence for this channel (Figure \ref{fig:events_hyp}, Panel D), suggesting that access to a community radio with a shared ethno-linguistic majority group is not sufficient to drive a change in behavior.\footnote{This is consistent with overall community radio having no effect: given the regional clustering of ethno-linguistic groups, same-language coverage correlates 0.80 with overall community radio coverage but only 0.50 with local community radio coverage (Table A2).}
	

		
	\subparagraph{Local community radio, interacted with ethno-linguistic belonging (Hypothesis 1c)}
	
	I also evaluate the role of shared ethno-linguistic belonging by examining whether local community radio has greater impacts in sub-prefectures widely sharing the ethnic makeup of their prefecture (and thus the radio station's main language), compared to others; as in equation (\ref{eq:flex_eb_lang}):\begin{align}
		y_{s,t} &= \sum_{\tau} \beta^{1}_\tau \; RadioShare^{Local}_{s} \times Ethnic^1_s \times \mathbbm{1}{_t(\tau)} \label{eq:flex_eb_lang} \\
		& + \sum_{\tau} \beta^{0}_\tau \; RadioShare^{Local}_{s} \times Ethnic^0_s \times \mathbbm{1}{_t(\tau)} + \textbf{X}_{s,t} \; \Gamma +\alpha_s +\lambda_t + u_{s,t} \nonumber
	\end{align}
	The dummy variable $Ethnic^1_s$ takes the value 1 if the fraction of sub-prefecture $s$ that speaks the language of the local community radio is above the median across sub-prefectures (vice versa for $Ethnic^0_s$).\footnote{From the 2014 census, I compute for each sub-prefecture the share of households whose language is the majority language of its prefecture (i.e. the main language of the local community radio): median 0.91, mean 0.77; OA Section D.3. $Ethnic^1_s=1$ ($Ethnic^0_s=1$) for the 171 (169) sub-prefectures above (at or below) the median, 112 (116) of which receive a local community radio signal.} If shared language or ethnicity makes local radio more effective, we expect a larger effect in sub-prefectures where the language of the local community radio is widely spoken ($\beta^{1}_\tau<\beta^{0}_\tau$, for $\tau>0$). The evidence for this is nuanced. Ebola infections fall after the scale-up of the campaign in both groups, with similar magnitude and timing, and the drop is statistically significant in both (Figure \ref{fig:events_hyp}, Panel E); the same holds for social resistance and treatment uptake (Figure A18).\footnote{Results for both groups are similar when $Ethnic^1_s$ is instead defined by the share speaking the local radio's language being above the mean; when it is instead defined by the majority language of the sub-prefecture coinciding with that of the prefecture, regardless of the share speaking it, the drop remains large in both groups but is imprecisely estimated in the small group (16\%) of sub-prefectures whose majority language differs (not reported).} A regression with the main effect of local radio and one interaction term with (demeaned) $Ethnic^1_s$, which estimates $\beta^{1}_\tau-\beta^{0}_\tau$ directly, shows relatively lower infections in sub-prefectures sharing the language of the local community radio in months 2 to 4 after the launch, suggestive of an earlier response, but no significant difference thereafter (Figure A19, Panel A), and no significant difference in any month when shared language is instead defined as at least 50\% of the population speaking the language of the local community radio (Panel B). Together with the evidence in Figure \ref{fig:events_hyp} Panel D, this indicates that, in this setting, the relevant community for achieving a change in behavior is defined by locality rather than by ethno-linguistic belonging.\footnote{It is possible that if there was zero ethno-linguistic overlap, with groups being perfectly segregated, that these would start to matter and add to the locality effect.}


	
	\begin{center}
		\begin{adjustwidth}{-1cm}{}
			\begin{figure}[H]
				\setstretch{1}\selectfont
				\center \caption{Event study: \textbf{Ebola infections} pre/post public health campaign \\
					by access to a given radio station \label{fig:events_hyp}}
				\subcaption*{A. Access to \textbf{any community radio} (H1) \hspace{50pt} B. Access to \textbf{national} radio (H2)}
											\vspace*{-.2cm}	
				\rotatebox{90}{\footnotesize \hspace{10pt} Log of Ebola cases per 100,000}
				\includegraphics[width=.46\columnwidth, trim=4 25 4 0,clip]{\figures/04_steps_lang/eventMdy_1RRT_\radio_leb01M_ci95_steps\bootrob.pdf} 
				\includegraphics[width=.46\columnwidth, trim=4 25 4 0,clip]{\figures/04_steps_lang/eventMdy_1RNT_\radio_leb01M_ci95_steps\bootrob.pdf} \\				
				\vspace*{.3cm}
				\subcaption*{C. Access to \textbf{local} community radio (H1a) \hspace{40pt} D. Access to a community radio \\
					\hspace*{220pt} 				\textbf{with shared majority language} (H1b) }	
											\vspace*{-.2cm}	
				\rotatebox{90}{\footnotesize \hspace{10pt} Log of Ebola cases per 100,000}
				\includegraphics[width=.46\columnwidth, trim=4 25 4 0,clip]{\figures/04_steps_lang/eventMdy_1RR_distT_\radio_leb01M_ci95_steps\bootrob.pdf} 					\includegraphics[width=.47\columnwidth, trim=4 25 4 0,clip]{\figures/04_steps_lang/eventlangMdy1_cov_RR_lang_\radio_leb01M_rr4_\bootrob_ci95.pdf}
				\vspace*{.3cm}
				\subcaption*{E. Access to \textbf{local} community radio, interacted with a dummy variable indicating whether a sub-prefecture shares the \textbf{majority language} of local community radio (left), or not (right) (H1c)}
				\subcaption*{ \hspace{20pt} $\times$ Shared Ethno-linguistic group \hspace{50pt} $\times$ Different Ethno-linguistic group}
				\vspace*{-.7cm}
				\center \rotatebox{90}{\footnotesize \hspace{10pt} Log of Ebola cases per 100,000}
				\includegraphics[width=.48\columnwidth, trim=4 25 4 0,clip]{\figures/16_steps_lang_app/eventlangMdy2_\radio_LANGdist_\lang_leb01M_rr4YES_\bootrob_ci95.pdf} 
				\includegraphics[width=.48\columnwidth, trim=4 25 4 0,clip]{\figures/16_steps_lang_app/eventlangMdy2_\radio_LANGdist_\lang_leb01M_rr4NO_\bootrob_ci95.pdf}
				\begin{tablenotes}
					\item \footnotesize \qquad \qquad With 95\% confidence intervals; bootstrap standard errors clustered by prefecture. 
					\vspace{0.1cm}
					\item[] \footnotesize \textit{Notes}: Coefficient estimates of equation (\ref{eq:flex_eb}) (and equation (\ref{eq:flex_eb_lang}) for Panel E), for different definitions of radio station, with omitted category -1 (May 2014). The outcome is the log-number of Ebola infections per 100,000 people +0.01. Interpreted as the percentage difference in Ebola infections for sub-prefectures with 1 pp. greater access to a given radio station, each month before and after the public health campaign was launched in all community radio stations. Because the outcome is in logs, differences in the coefficient path over time correspond to differences in the growth of infections. The vertical lines indicate the launch of the campaign (gray dashed line), and the scale-up of the campaign as significant international aid arrived (red solid line). All panels are conditional on controls, which include log-population, geographic area, distance to epicenter, and distance to the closest community radio transmitter, interacted with time-dummies, and distance to the closest ETU, CCC and laboratory interacted with time-dummies. Panels C-E also condition on the share of the sub-prefecture covered by any community radio station, interacted with time-dummies. Estimated coefficients are in Tables C2-C4. Figure~A10 also conditions on overall radio access.
				\end{tablenotes}
			\end{figure}
		\end{adjustwidth}
	\end{center}

	\subparagraph{Understanding the roles of geographic locality and ethnic-belonging.} 
	
	Several pieces of evidence can help us understand why geographic locality, more than ethnic identification, drives the impacts of local community radio on Ebola. Guinea, as many African countries, exhibits diversity both within and across regions and locations, so there is ample opportunity for interaction between groups. When evaluating the role of a prefecture as the relevant community, one needs to consider that administrative divisions were to a certain extent imposed by the French colonizing power, but their origins predate the colonial period \citep{goerg2011couper, mcgovern2012unmasking}. They relate to ethno-linguistic groups, but a given prefecture and sub-prefecture still harbors important diversity |suggesting that a community is not solely defined by ethnicity. Panel A in Table A5, which presents ethnic and religious fractionalization across the country, regions, prefectures and sub-prefectures, illustrates this. The probability that two people belong to different ethnic groups is approximately 0.27 (0.16) at the prefecture (sub-prefecture) level, compared to 0.74 (0.43) at the country (regional) level. This suggests that people are in close contact with other ethnicities, even within a prefecture or sub-prefecture. It also suggests that prefectures are only somewhat more diverse than sub-prefectures, and that both have more homogeneous groupings, compared to regions or the country. This point is also illustrated by the fact that 76\% (83\%) of individuals in a prefecture (sub-prefecture) speak that unit's majority language, and 24\% (17\%) do not (Panel B of Table A5 and Figure C7). Further, Guineans speak an average of at least 2, and up to 5, languages (Panel C of Table A5). Finally, community radio stations take existing diversity into account and transmit radio programs in a number of languages. During the study period, stations report airing mainly in one or two languages, which coincide empirically with their prefectures' majority groups (OA Section F).\footnote{In addition to the main language, they may use French and other local languages. E.g. the community radio of Boké reports using mainly Susu and Fulla, but also sometimes Landuma, Diakanke, Nalou, Mikhifore and French.}

\paragraph{Summary: mapping results to hypothesized channels.} Together, the results described are consistent with the model predictions (Section~\ref{sec:concept}): I find no evidence to support \textit{Hypothesis~1} and \textit{Hypothesis~2} (any community radio, national radio), arguing against a generic ``any radio'' channel. I find no support for \textit{Hypothesis~1b}, and limited support for \textit{Hypothesis~1c}: sub-prefectures with greater ethno-linguistic overlap with the local community radio show a suggestive earlier response, but the local radio is effective in both groups. This suggests ethno-linguistic identity is not sufficient (\textit{Hypothesis~1b}) to determine the relevant ``other'' in this setting. And although shared ethnic makeup may somewhat complement locality, high ethno-linguistic overlap is not necessary for the local radio to be effective (\textit{Hypothesis~1c}). Instead I find evidence in favor of \textit{Hypothesis~1a} (on local community radio), consistent with \textit{locality} defining the relevant ``other'' in this setting. Section~\ref{sec:mech} returns to whether this locality advantage operates through norm proximity or through informational precision/trust.

	\subsection{Examining the impacts of local community radio (Hypothesis 1a) \label{sec:event_local_eb}}
	
	Having ruled out competing hypotheses, I examine further the main hypothesis, that local community radio helped curb the spread of Ebola. Figure \ref{fig:events_local} presents results of the impact of local community radio on the spread of Ebola and on social health behavior (discussed next in Section \ref{sec:mech}), with and without covariates, and controlling for overall community-radio access.

	Results are similar whether examining the impact of local community radio on the spread of Ebola, unconditionally (Figure \ref{fig:events_hyp}, Panel C), or conditional on overall community-radio access (Figure \ref{fig:events_local}, Panel A).
	First, they support the parallel-trends assumption. I find no difference in the growth of Ebola infections in areas with varying levels of access to a local community radio station prior to the launch of the public health campaign, and at least until its effective adoption by community radio stations throughout the country in September 2014. 

	Second, the public health campaign leads to a drop in Ebola infections about seven months after the intended starting date, or four months since its wide adoption by most community radio stations. While this appears as a long time lag, results are interpreted as intention-to-treat effects, given that the effective and wide adoption of the sensitization campaign took several months after its official launch. An alternative strategy using the actual start date of the sensitization campaigns when available shows a much more immediate effect (Section \ref{sec:events_dcdh}).
	
		Third, the results indicate that a $10$ percentage point ($pp$) increase in access to a local radio station (from an average of 27\% coverage) lowers Ebola infections per 100,000 people by $21\%$ in the fourth month since the scale-up of the campaign, or by an average of $\sim$17\% per month since that date (also in Table C5). Effects subside a year into the campaign, in June 2015, as the epidemic is coming to an end due to the combination of public health measures put in place.\footnote{Partly this can be due to new sensitization campaigns in the second half of 2015 (Table A6), and new local radio stations, but only one is launched in March 2015, two in May 2015, the others earliest in August 2015.}
			
		A back-of-the-envelope calculation, focusing on places with above-median access to community radio, suggests that around 664 infections, or 17\% of total cases, could have been prevented if places with above-median access to some community radio had had their own. This is computed by multiplying each event-study estimate since the announced launch of the campaign, with the corresponding number of new Ebola cases in the control group |places with above-median access to community radio, but without one of their own.\footnote{Alternatively, taking the average of the coefficients over the period since we start seeing statistically significant effects ($\tau=+7$), or since the announced launch of the campaign ($\tau=0$) until the last event study effect ($\tau=+12$), and multiplying either by the total Ebola burden during the corresponding months in the control group, implies a reduction by 447 cases, a 12\% reduction, or by 357 cases, a 9\% reduction, respectively.} 
		

\begin{center}
	\begin{adjustwidth}{-5cm}{}
		\begin{figure}[H]
			\setstretch{1}\selectfont
			\vspace{-.5cm}
			\center \caption{Event study: \textbf{Ebola infections}, \textbf{social resistance} and \textbf{treatment-uptake} \\
				pre/post public health campaign, by access to a \textbf{local} community radio station (H1a), \\
				conditional on overall community-radio access \label{fig:events_local}}
			\vspace*{-.2cm}
			\foreach \var/\ylabelA/\ylabelB/\title in {
				lebola01M/{\hspace{10pt} Log of Ebola cases}/{\hspace{10pt} per 100,000 people}/{A. Ebola Infections},
				rsouspref01_NEW/{Weeks per month with}/{Social resistance events}/{B. Social resistance},
				ebd_CE_M_pc/{Ebola deaths in community over}/{total deaths, per 100,000 people}/{C. Treatment-uptake: ratio of Ebola deaths in-community over total deaths}%
			}{
				\subcaption*{\title}
				\vspace*{-.2cm}
				\rotatebox{90}{\footnotesize \hspace{20pt} \ylabelA}
				\rotatebox{90}{\footnotesize \hspace{20pt} \ylabelB}
				\includegraphics[width=.46\columnwidth, trim=4 25 4 0,clip]{\figures/06_event_main/eventMdy_RR_dist_\radio_\var_rr1_yonah_\bootrob_ci95.pdf}
				\includegraphics[width=.46\columnwidth, trim=4 25 4 0,clip]{\figures/06_event_main/eventMdy_RR_dist_\radio_\var_rr4_yonah_\bootrob_ci95.pdf} \\
				\hspace{10pt} \footnotesize{Without controls} \hspace{150pt} \footnotesize{With controls} \\
			}
			\begin{tablenotes}
				\item \footnotesize \qquad With 95\% confidence intervals; bootstrap standard errors clustered by prefecture; Omitted category: t-1. 
				\vspace{0.1cm}
				\item[] \footnotesize \textit{Notes}: Coefficient estimates of equation (\ref{eq:flex_eb}), with omitted category -1 (May 2014). Coefficients are interpreted, for sub-prefectures with 1 pp. greater access to a local community radio station, as the percentage difference in Ebola infections per 100,000 people (panel A); divided by 100, as the difference in the number of weeks per month with social resistance (panel B) or in treatment uptake (panel C), each month before and after the public health campaign was launched. The vertical lines indicate the launch of the campaign (gray dashed line), and the scale-up of the campaign as significant international aid arrived (red solid line). I condition on access to community radio. Controls include log-population, geographic area, distance to epicenter, and distance to the closest community radio transmitter, interacted with time-dummies, and distance to the closest ETU, CCC and laboratory interacted with time-dummies. Social resistance is measured as the number of weeks per month with civil violence or opposition to public health measures. Treatment uptake is the number of Ebola deaths in-community over total Ebola deaths + 1, both per 100,000 people. Figures A13-A14 give alternative definitions. Estimated coefficients are in Table C5.
				\vspace{-.5cm}
			\end{tablenotes}
		\end{figure}
	\end{adjustwidth}
\end{center}

\subsection{Actual timing of sensitization campaign}\label{sec:events_dcdh}	

A caveat of the DiD strategy above, using the intended starting date of the sensitization campaign, is that community radio stations gradually adopted the guidelines from the headquarters of RRG in Conakry. The estimated effects take months to materialize, raising the potential concern that something else caused the drop in Ebola infections, other than the information campaign itself. I provide an alternative strategy to validate the previous results, using the actual starting date of the sensitization campaign across different community radio stations. 

The actual date in which different community radio stations started the sensitization campaign is available for 11 out of the 22 prefectures with a pre-existing community radio station. 9 of them launched the full-fledged sensitization campaign after September 2014.\footnote{The first is Guéckédou in March 2014, followed by Kouroussa in June 2014; the other 9 started between September and December 2014 (see details in OA Sections D.2 and F). New community radio stations opened between April 2015 and March 2016 (footnote \ref{foot:newradio}).} 

This set-up with more than two time periods and variation in the timing of when different locations start the public health campaign, diverges from the canonical DiD design. To account for potentially heterogeneous treatment effects I apply the estimator proposed by \citet{de2024difference} (DCDH) to equation (\ref{eq:flex_eb}) above, where the post-treatment dummy now varies by radio station.\footnote{$\mathbbm{1}{_{s,t}(\tau)}$ has subscript $s$ for sub-prefecture, $t$ for time. It differs by sub-prefecture, according to the start date of the sensitization campaign in the radio station that is local to them, i.e. that of the prefecture it belongs to.} 	Since I only observe the starting date of the sensitization campaign for a fraction of places, I estimate equation (\ref{eq:flex_eb}) with DCDH using the 18 prefectures for which this date is observed (11 pre-existing and 7 new stations).\footnote{I use the Stata command \texttt{did\_multiplegt\_dyn} \citep{de2025stata}, comparing treated to not-yet-treated units. Prefectures whose station has no observed start date are excluded from estimation by construction, whether the station predates the outbreak or not.} 

The identifying assumption is again parallel trends, which I evaluate in pre-campaign Ebola dynamics. It is possible, however, that the timing of effective adoption is correlated with the spread of Ebola. I find no correlation between adoption timing and the total or cumulative number of Ebola cases at the time of adoption. Later adoption is associated with greater distance to the epicenter, suggesting that selection into early treatment occurs along a geographic dimension. My baseline specification controls for distance to the epicenter with a time-varying effect, which absorbs this dimension of selection.

	Results presented in Panel A Figure \ref{fig:events_dcdh_chg} (and Table C7) show that the sensitization campaign on local radio led to a drop in Ebola infections, in the ratio of people dying in the community over total Ebola deaths, and in social resistance. The dynamics are consistent with a causal interpretation: estimates are flat and close to zero in the six months before a station adopts the campaign and in the adoption month itself, break downward in the second month, and then decline steadily. The estimated effect of 10pp greater access to local community radio, relative to not-yet-treated units, is a reduction in Ebola infections of about $6\%$ in the second month after a station adopts the campaign, up to $11\%$ by the fifth month and $16\%$ by the twelfth, consistent with the cumulative nature of exposure to the campaign; the effects are jointly significant ($p=0.02$), as is the average total effect (5\% level).


\begin{center}
	\begin{adjustwidth}{-5cm}{}
		\begin{figure}[H]
			\setstretch{1}\selectfont
			\vspace{-.9cm}
			\center \caption{Event study: \textbf{Ebola infections}, \textbf{social resistance} and \textbf{treatment-uptake} \\
				pre/post public health campaign, by access to a \textbf{local} community radio station (H1a), \\
				conditional on overall community-radio access \\
				| Actual start of radio campaign \label{fig:events_dcdh_chg}}
			\vspace*{-.2cm}
			\foreach \var/\ylabelA/\ylabelB/\title in {
				lebola01M/{\hspace{30pt} Log of Ebola cases}/{\hspace{30pt} per 100,000 people}/{A. Ebola Infections},
				rsouspref01_NEW/{\hspace{30pt} Weeks per month with}/{\hspace{30pt} Social resistance events}/{B. Social resistance},
				ebd_CE_M_pc/{\hspace{5pt} Deaths in community over}/{\hspace{5pt} total deaths per 100,000 people}/{C. Treatment-uptake: ratio of Ebola deaths in-community over total deaths}
			}{
				\subcaption*{\title}
				\vspace*{-.2cm}
				\rotatebox{90}{\footnotesize \ylabelA}
				\rotatebox{90}{\footnotesize \ylabelB}
				\includegraphics[width=.47\columnwidth, trim=4 25 4 40,clip]{\figures/05_did_endog/radio_chg_\var_ctrl1.pdf}
				\includegraphics[width=.47\columnwidth, trim=4 25 4 40,clip]{\figures/05_did_endog/radio_chg_\var_ctrl4.pdf} \\
				\vspace*{-.4cm}
				{\footnotesize \hspace{20pt} Without controls \hspace{140pt} With controls}\\		
				\vspace*{.2cm}	
			}
			\begin{tablenotes}
				\item \footnotesize \qquad With 95\% (light shade) and 90\% (darker shade) confidence intervals; robust standard errors clustered by prefecture (analytic); omitted category: t-1. \\
				\quad \\
				\item[] \footnotesize \textit{Notes}: Event study of the effects of the sensitization campaign using the \citet{de2024difference} (DCDH) estimator, comparing treated to not-yet-treated sub-prefectures, with event time defined relative to each community radio station's actual adoption of the campaign, for the 18 prefectures with an observed adoption date, and the 7 prefectures whose first local station opened during the outbreak are treated from that station's start date, with the new station's coverage as treatment dose. Coefficients are interpreted, for sub-prefectures with 1 pp. greater access to a local community radio station, as the percentage difference in Ebola infections per 100,000 people (panel A); divided by 100, as the difference in the number of weeks per month with social resistance events (panel B) or in the ratio of Ebola deaths in-community over total deaths + 1, both per 100,000 people (panel C), in each month relative to the station's adoption of the campaign. I condition on access to community radio. Controls include log-population, geographic area, distance to the epicenter, distance to the closest community radio transmitter, and distance to the closest ETU, CCC and laboratory, each interacted with a monthly time trend. Estimated coefficients are reported in Table C7; alternative definitions of the control group in Tables C8- C9. 
				\vspace{-.5cm}
			\end{tablenotes}
		\end{figure}
	\end{adjustwidth}
\end{center}

\paragraph{Interpreting lagged DiD estimates.}\label{sec:interpret}

Together these findings suggest that the effects of the sensitization campaign are quite immediate, about two months since the start of the campaign. This contrasts with the seven or four months lag found with the most conservative approach of using either the official launch of the public health campaign by the director of the community radio stations as reference period, or a somewhat less conservative approach of using the date when the non-governmental organization \textit{Fondation Hirondelle} provided technical assistance and content in multiple local languages for the community radio stations to upload. This ensures that the pre-treatment period offers a clean comparison, but also overestimates the time-lag of the campaign. Given that radio stations actually adopted the sensitization campaign later, I interpret the DiD results as intention-to-treat effects.
	
Some lag in the effect of information is consistent with a mechanism of coordination in behaviors attached to social norms. Game theory insights predict that it could take time to adjust behaviors that depend on others' beliefs and actions, based on changes in information that are common knowledge \citep{geanakoplos1992common}. While epidemiological models consider the role of information on shortening epidemics \citep{funk2009spread, kiss2010impact}, less is known about how fast public information actually spreads and impacts epidemics. Empirical evidence shows that individual health behavior, such as choosing whether to travel in the next days during the Covid-19 pandemic or not, can change rapidly due to targeted advertising on social media \citep{breza2021effects}. Socially-entrenched behaviors, on the other hand, may take more time to change. For example, \citet{wakefield2011effects} find that a media campaign impacted smoking behavior three months after the campaign started. Public health literature highlights the need for widespread, frequent and prolonged exposure for a campaign's success \citep[e.g.][]{flay1987mass, hornik2002public, wakefield2010use}. Communications research also advises that most campaigns take anywhere from 90 days to a year to complete \citep{blakeman2011advertising}.

	\subsection{Robustness checks \label{sec:robust}}

\paragraph{Assessing potential confounders.}
	
A potential concern is that the epidemic spreads faster depending on underlying socio-economic characteristics correlated with access to local radio, that are activated only after some time, for example, due to the arrival of aid. 

\textit{Socio-economic observables.} To address this, I examine whether access to local radio is correlated with key socio-economic characteristics. I find that overall access to radio is correlated with important socio-economic observables (columns 1-5 in Table A8). However, access to local radio, conditional on access to overall community radio, is not systematically related to socio-economic characteristics predictive of the spread of disease (columns 6-7 in Table A8).\footnote{Trust in president appears borderline significant in column 6; addressed below under alternative mechanisms.}

\textit{Geography and disease.} There is some evidence that larger territories have greater access to local radio (column 5 in Table A9), but not after controlling for community radio overall (column 6-7). All specifications include controls for log-population, geographic area and distance to the epicenter interacted with time dummies. Reassuringly, I also find no evidence that other diseases, such as cholera, malaria, yellow fever or the flu are systematically related to access to local radio prior to the epidemic that led to the public health campaign (Table A10).\footnote{This data is only available at prefecture-level.}

\textit{Unobservables.} To ensure that impacts are not driven by unobserved characteristics that fundamentally differentiate prefectures that have a local radio station from those that do not, I do three exercises, which confirm the baseline results. First, I exclude sub-prefectures where the community radio transmitter is located (panel A, Figure A15). Second, I control for the location of the community radio transmitter by including a dummy variable indicating whether the sub-prefecture has a community radio transmitter or not, interacted with each event-study time dummy (panel B, Figure A15). Third, prefectures, as opposed to sub-prefectures, with a radio transmitter could be different to begin with. To address this I confirm that results are robust to controlling for a prefecture's access to any local community radio, that is, they also hold at the intensive margin of greater access to a local community radio (panel C, Figure A15). 

\paragraph{Assessing alternative explanations.}

\quad \\
\indent \textit{Government accountability.}	Local media could affect the epidemic through a different mechanism, due to its impact on the provision of public goods. Better access to information can lead governments to provide more public goods because citizens are more able to hold them accountable \citep{besley2002political, eisensee2007news}. The most relevant public policy and the best proxy for national and international efforts to contain the epidemic is the establishment of health facilities, namely Ebola treatment units (ETUs), laboratories for rapid testing, and pre-existing Community care centers (CCCs). The evidence does not support this channel. First, results are robust to the inclusion of time-varying access to these public health facilities as controls. Second, places with greater access to a local radio station are not more likely to open ETUs, laboratories or have CCCs, whether measured by the distance to the closest public health facility, or the number of them adjusted by population (panels A and B, columns 5-7, Table A11). Third, I study whether public health measures saw a similar change around the time at which we observe impacts on Ebola, and whether this differentially affected places with distinct exposure to local radio stations. Figure A9 provides evidence against this.

\textit{Trust as confounder.}	Public health authorities that are trusted may be better able to affect health behavior and hence the epidemic. If trust in state institutions is correlated with pre-existing access to radio stations, and it is activated at a later stage of the disease, it may explain the results. Trust in president is borderline (column 6), but not robustly correlated with access to local community radio after controlling for overall community radio (column 7, Table A8). Results replicate when controlling for trust in president, interacted with time dummies, Figure A16, indicating that local radio is not merely capturing pre-existing differences in trust. Section \ref{sec:mech_trust} examines trust in media as a mechanism of the local radio effect, rather than a confounder.
	
	\paragraph{Alternative specifications.}
	\quad \\
	\indent \textit{Lagged dependent variable.} Results replicate when conditioning on Ebola lagged one period, instead of fixed effects (panel C, Figure A11). They are a lower bound estimate of the true effect of local radio \citep{angrist2008mostly}.
	

\textit{Alternative outcomes.} Results are robust to alternative measures of the outcome variable. The baseline specification has $log(Ebola+0.01)$ as the dependent variable, to approximate the epidemiological theory. Results are identical in terms of the timing of the drop in infections, and of similar magnitude, when using alternative outcome definitions, i.e. $log(Ebola+1)$, the number of Ebola cases, or adjusted by population (Figure A11), and measures, i.e. using confirmed Ebola only, excluding probable cases, or focusing on Ebola deaths (Figure A12).

\textit{Alternative signal-reception threshold.} The baseline radio-coverage measures are based on a signal strength of at least 43 dB $\mu$V/m, required for normal receivers. Results are robust to re-constructing all coverage measures with the lower 33 dB $\mu$V/m threshold (raw data in Figure C2, Panels C--D; replication results in Figure C8 and Table C6).\footnote{These are the two standard thresholds provided by the Nautel Radio Coverage Tool online software used to compute radio signal reception.} 

\textit{Alternative controls.} Results also hold with more parsimonious specifications, not conditioning on other radio nor controls (Figure A17 adds radio signal and controls separately).

\textit{Alternative comparisons.} DCDH results are qualitatively similar when considering only pre-existing stations as treated, and the rest as never-treated, either including in the control group only those with a known opening date (Table C8), or instead using all of them (Table C9).
		
	\section{Mechanisms \label{sec:mech}}
	
Having shown that local community radio impacted Ebola infections (Section~\ref{sec:empirical}), this section asks \textit{why}. The framework in Section~\ref{sec:concept} yields two predictions tested here. First, public information shifts socially-sanctioned behaviors more than private information does, while the public--private distinction matters little for behaviors that are private or not reprimanded |and may even reverse when positive externalities induce free-riding. Second, locality may matter through three channels: contagion proximity, informational precision and trust, and norm proximity. \textit{Contagion proximity} is common across signals for a given listener and therefore does not differentiate local from non-local radio. The earlier rejection of overall community radio (H1) rules out this channel, since any radio with the same content should produce similar effects. This section explores the remaining two channels. The remaining two yield a sharp empirical test: under \textit{norm proximity}, local radio should facilitate change in socially-sanctioned but \emph{not} in private behaviors, relative to other sources of information, whereas under \textit{informational precision or trust} it should have an advantage for both.

I proceed in four steps. First, using the same DiD design as above, I evaluate the impacts of local community radio on two socially-sanctioned outcomes: social resistance and treatment uptake (Section~\ref{sec:mech_social}). Second, I examine whether public and private information differentially affect safe burial refusals (a socially-sanctioned outcome), evaluating two types of Red Cross campaigns: local mobile-radio (public) and door-to-door social mobilization (private) (Section~\ref{sec:mech_ifrc}). Third, I use cross-sectional survey data to compare the effects of media on second-order beliefs vs. private behaviors, distinguishing norm proximity from informational precision and trust (Section~\ref{sec:mech_beliefs}). Fourth, I evaluate trust specifically, using heterogeneity by pre-existing trust in radio and stated preferences for community radio (Section~\ref{sec:mech_trust}).

	\subsection{Local radio impacts on socially-sanctioned behavior \label{sec:mech_social}}

Using the same DiD strategy as above (equation~\ref{eq:flex_eb}), I evaluate the impacts of local community radio on two socially-sanctioned outcomes: social resistance and treatment uptake. Results are also validated with variation in the actual start of the sensitization campaign (Section \ref{sec:events_dcdh}).

	First, I study impacts on social resistance events, which are collective acts expressing a refusal to abide by public health measures (see Section \ref{sec:data_out}). There is a contemporaneous drop in social resistance in the same month as the drop in the transmission rate of Ebola (Panel B, Figure \ref{fig:events_local}). This is consistent with social resistance behavior and the spread of the virus reinforcing each other. Most of the impacts arise four months since the campaign's wide adoption, starting from a $53\%$ drop in social resistance in the fourth month, and an average of $25\%$ per month since then, for $10pp$ greater access to a local radio station. Using the actual start of the radio campaign when this date is available (Panel B, Figure \ref{fig:events_dcdh_chg} and Table C7), gives a similar same picture: social resistance falls two months after the actual start of the campaign, although this effect is only statistically significant a few months later. $10pp$ greater access to local radio leads to about $0.03$ fewer weeks with resistance events per month ($32\%$ of the sample mean) in the fifth and sixth months, and persistent declines thereafter that are jointly significant ($p=0.03$).
	
	Second, I examine impacts on treatment-uptake, measured as the ratio of Ebola deaths in the community over the total number of Ebola deaths, including those that die in an ETU.\footnote{The outcome is the number of Ebola deaths in-community per 100,000 people over total Ebola deaths per 100,000 people + 1, times 100 (see OA Section D); results are robust to other definitions} I find a gradual decrease in the relative number of Ebola deaths in-community, by $31\%$ in the fourth month, although noisily estimated, and an average of $20\%$ per month since then, for $10pp$ greater access to a local radio station. This is confirmed using the actual timing of the campaign (Panel C, Figure \ref{fig:events_dcdh_chg} and Table C7): the ratio of Ebola deaths in-community falls by $0.6$--$0.9$ per 100,000 per month ($30$--$48\%$ of the sample mean) from the second month onward, significant at the 5\% level in nearly every month ($0.5$--$0.8$ percentage points in levels).
	
	The impacts of local radio are not driven by ethno-linguistic belonging also for these outcomes: social resistance and treatment uptake fall regardless of the ethno-linguistic majority of sub-prefectures (Figure A18). Together these results show that local radio impacted socially-sanctioned behavior |in line with the model predictions|, and are consistent with this behavior change contributing to the drop in Ebola cases.

	

	\subsection{Public vs private local information impacts on cultural practices \label{sec:mech_ifrc}}

Having established that local radio impacted socially-sanctioned behavior, I test whether public information has a larger impact than private information on this type of behavior, by exploiting within-Red-Cross variation between local radio campaigns (public information) and door-to-door social mobilization (private information) on safe burial refusals. 
	
To accompany safe and dignified burials (SDBs), the Red Cross broadcast radio spots on community radio, went through communities airing mobile radio shows from their vehicles, and conducted door-to-door social mobilization campaigns. These campaigns explained the value of SDBs, and reported how communities were gradually adopting them. Since this data is only available from April 2015, late into the epidemic and months after community radio broadcast the sensitization campaign, I am not able to apply the difference-in-differences strategy used above. Instead, I exploit variation per prefecture per month in the intensity of mobile radio shows, on one hand, and in door-to-door campaigns that are classified by the Red Cross as preventive, on the other |as opposed to other campaigns conducted in response to attacks.  
	
I conduct an event study analysis, evaluating the effect of mobile radio and preventive door-to-door campaigns conducted by the Red Cross across different prefectures on SDB refusals over time, conditional on total number of SDBs conducted by the Red Cross that month. Since different prefectures are treated with varying levels of intensity over time, I estimate equation (\ref{eq:burial1}) using the estimator developed by \citet{de2024difference} (DCDH).\begin{align} \label{eq:burial1} 
	refuse_{s,t} &= \sum_{\tau} \beta_\tau radio_{s,t-\tau} + \sum_{\tau} \gamma_\tau door_{s,t-\tau} + \delta SDBs_{s,t} + \textbf{X}_{s,t} \Gamma +\alpha_s +\lambda_t + u_{s,t} 
\end{align}

	The outcome variable $refuse_{s,t}$ counts the number of SDB refusals in prefecture $s$ at month $t$. The treatment variables of interest are $radio_{s,t}$, an indicator variable for whether the Red Cross aired a mobile radio campaign, and $door_{s,t}$, the number of preemptive social mobilization campaigns in prefecture $s$ at month $t$. I control for the total number of SDBs in the same time period. $\bf X_{s,t}$ is a vector of controls with indicators for epidemic surveillance efforts by the Red Cross in the same period (number of contacts followed, number of sick transported to ETUs, and chlorine available), as well as public health efforts (distance to the closest ETU, laboratory and CCC) and log-population interacted with a monthly time-trend.\footnote{Results are not sensitive to excluding these controls. Distance to epicenter and geographic area are not included here since at the prefecture-level they are near-collinear with the other distance controls.} Time-invariant differences across prefectures and time-fixed effects of unknown form are captured by $\alpha_s$ and $\lambda_t$. Standard errors are clustered at prefecture-level, to account for serial dependence.
	
		The coefficients of interest are $\{\beta_\tau\}_{\tau\geq 0}$ (or $\{\gamma_\tau\}_{\tau\geq 0}$). They measure the impact of a mobile-radio campaign (or an increase in door-to-door campaigns) on refused SDBs after $\tau$ months. This specification allows me to test the parallel-trends assumption, by evaluating whether $\beta_\tau$ ($\gamma_\tau$) is constant for all $\tau<0$, i.e. before a given mobile-radio (door-to-door) campaign.
			
The social mobilization efforts were likely endogenous to the spread of Ebola and to the change in health behavior over time. In particular, communities that were more intensely impacted by the virus or that were more likely to refuse safe burials or treatment, are plausibly more likely to be targeted with these campaigns. This, however, would lead to a positive spurious correlation, biasing us against finding that these campaigns reduce SDB refusals. Arguably, the Red Cross would substitute door-to-door campaigns with mobile radio in places where they expect more violence and contact with the virus. This would make it more difficult to detect an effect of mobile radio on refused SDBs if it was there. To minimize these concerns, I control for the total number of SDBs and other epidemic surveillance efforts that month.

The identifying assumption is that in the absence of these social mobilization campaigns, refusals of SDBs would have evolved in parallel, conditional on the total number of SDBs and other epidemic surveillance efforts that month. The event study analysis also further allows me to evaluate the existence of pre-trends.

	\subsubsection*{Results on the impact of private vs public information on cultural practices}
	
\begin{figure}[hb!]
		\setstretch{1}\selectfont
		\caption{Refused ``Safe and Dignified Burials'' (SDBs) by IFRC Campaign \label{fig:ifrc_dcdh}} 
		\centering
		\subcaptionbox{Effect of mobile-radio campaigns \\ on refused SDBs 
			\vspace*{-.2cm} \label{fig:ifrc_dcdhA}}[0.45\textwidth]{%
			\includegraphics[width=.5\columnwidth, trim=0 0 0 20,clip]{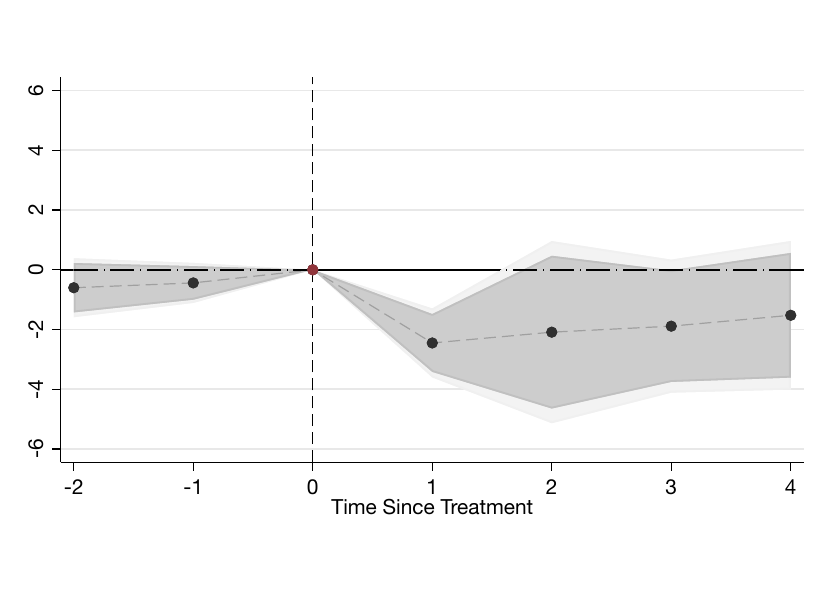}		
		}
		\hfill
		\hspace*{-1cm}\subcaptionbox{Effect of door-to-door campaigns \\ on refused SDBs 			\vspace*{-.2cm} \label{fig:ifrc_dcdhB}}[0.45\textwidth]{%
			\hspace*{-1cm}\includegraphics[width=.5\columnwidth, trim=0 0 0 20,clip]{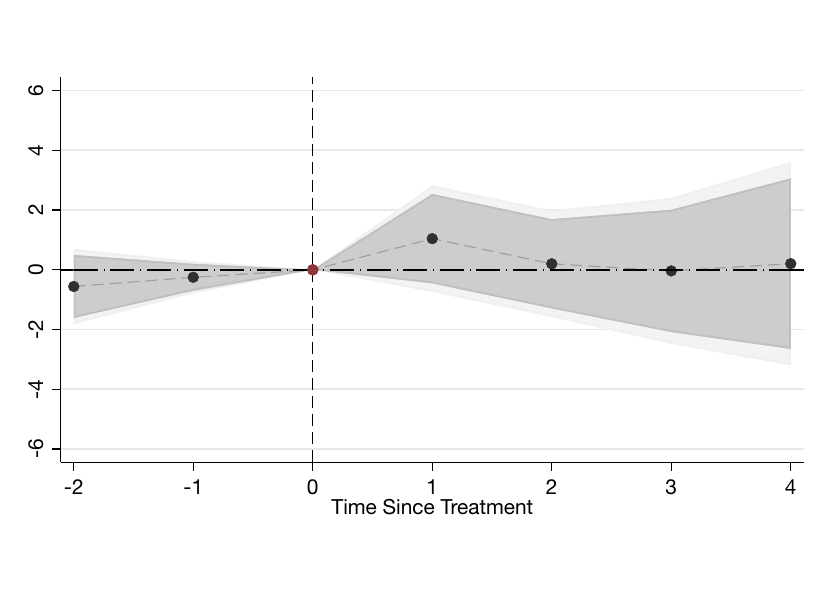}	
		}
		\begin{tablenotes}
				\item \footnotesize \quad With 95\% (90\%) [light shade (darker)] confidence intervals; robust standard errors clustered by prefecture. 
\vspace{0.1cm}			
			\item[] \footnotesize \textit{Notes}: Impacts of IFRC mobile-radio emissions (left), or preventive door-to-door campaigns (right) on number of refused ``safe and dignified burials'' (SDBs) in prefecture, conditional on the total number of SDBs in the same period, and on prefecture and time fixed effects. Estimated using \citet{de2024difference} (DCDH), comparing treated to not-yet-treated prefectures. It controls for epidemic surveillance efforts in the same period (number of contacts followed, number of sick transported to ETUs, and chlorine available), public health efforts (distance to the closest ETU, laboratory and CCC) and log-population interacted with a monthly time-trend. Estimated coefficients, and alternative specifications are reported in Table C10.
		\end{tablenotes}
	\end{figure}

	The results in Figure \ref{fig:ifrc_dcdh} show no evidence of pre-trends in refused SDBs prior to a new mobile-radio show aired in a given prefecture in a given month, nor prior to an increase in preventive door-to-door campaigns (with additional results in Table C10). In the month following the airing of mobile-radio shows, we observe a significant and sizable reduction in refusals of SDBs.
	 Mobile-radio leads to an average of 2.6 fewer refused burials per month in a given prefecture, compared to months without mobile radio.\footnote{Relative to an average (maximum) of .34 (8) refused SDBs per month per prefecture (Table A3). A 1 standard-deviation (SD) increase in the incidence of mobile-radio leads to a reduction by .75 of a SD in SDBs.} On the other hand, we do not see a drop in refused SDBs following an increase in door-to-door campaigns. If anything the coefficient appears to be positive, but not statistically significant. These findings are consistent with the conceptual framework above: public information at the local level, here via mobile-radio shows, helps achieve a coordinated change in community norms around burials. Private information, such as door-to-door campaigns, on the other hand is not sufficient to achieve this change. While a door-to-door campaign can be more personal than a mobile-radio show, it does not achieve the common knowledge effect required to shift norms |one does not know whether others received the same information as oneself.

	\subsection{Socially-sanctioned vs private health behavior \label{sec:mech_beliefs}}

		To disentangle the role of \emph{norm proximity} from \emph{informational precision or trust} in explaining the impacts of local media, I evaluate the effects of media on second order beliefs or behaviors that are socially-sanctioned on one hand, and on private behaviors on the other. Under norm proximity, local radio should affect socially-sanctioned behavior but not private behavior; under informational precision or trust, it should affect both (Section \ref{sec:concept_dist}).
		
	To do so, I study whether learning about Ebola from the media, which is predominantly radio, affects two different types of behavior.\footnote{The media source is predominantly radio, but it may include TV in a few cases. Other means of information include door-to-door campaigns, information by an NGO or by a doctor.} First, the belief that neighbors seek treatment when sick. This is a proxy for second-order beliefs about going to ETUs, which is socially-sanctioned (Section \ref{sec:data_out}). Second, the use of chlorine for hand-washing, clothes and other personal items. This behavior is either private or not socially-costly. Using cross-sectional survey data collected towards the end of the outbreak (described in Section \ref{sec:data_out}), I evaluate the effect of having learned about Ebola on the media, conditional on having received any information about Ebola, and demographic controls, estimating equation (\ref{eq:info_impacts1}): \begin{align}
		& belief_i = \alpha + \beta \; Media_{i} + \gamma \; EbolaInfo_{i} + \textbf{X}_{i,s} \; \Gamma + u_{i,s} \label{eq:info_impacts1}
	\end{align}
	
	The outcome is the belief or behavior reported by individual $i$. I condition on $EbolaInfo_{i}$, which is a dummy variable indicating whether an individual reports having received any information about Ebola.

	I take different approaches to define treatment $ Media_{i} $, and estimate the above equation (\ref{eq:info_impacts1}): 
	
	(1) I define $ Media_{i} $ as a dummy variable indicating whether individual $i$ has learned about Ebola from the media. To address the concern that people who seek health information are also more likely to know more about Ebola and take protective measures, I verify results instrumenting this variable with local community radio coverage. The exclusion restriction is that local community radio coverage does not drive the observed behavior other than through its effect on learning about Ebola from the media. My preferred specification conditions on comparable places |those with above-median access to radio; results are replicated with the full sample.
	
	(2) Alternatively, I define $ Media_{i} $ as the proportion $k$ of people in one's own sub-prefecture who learned about Ebola from the media, except oneself. I also verify results using the same IV strategy to address the concern that neighbors' information-seeking is also a function of their willingness to take protective actions. I use the first stage regressions to evaluate whether local community radio has greater predictive power on the own media-exposure dummy, or on the leave-one-out share of others with media-exposure |suggesting coordination. 
	
		(3) To uncover the common-knowledge aspect of media, I examine the presence of non-linearities in $k$. If coordination between people is important, effects should be larger as more people are informed. I include three dummy variables $Media_{i}^k$ $\forall k$, indicating whether the share of people in one's own sub-prefecture (excluding oneself) who learned about Ebola from media is in the bottom-, mid-, and top-tercile above-zero; the omitted category is sub-prefectures where no one else learned about Ebola from the media. I include similar dummies for different shares of individuals in one's sub-prefecture that have received any information about Ebola.
	
	$\textbf{X}_{i,s}$ controls for individual and household characteristics.\footnote{I include all baseline controls available in the survey, namely age, gender, nationality, education, marital status, employment status, household size, number of household members working, number of students, number of children under five, drinking-water and lighting source, housing tenure and type, number of bedrooms, roof and floor materials, and a wealth index. I include log-population and geographic area.} In addition, to capture the role of local community radio, as opposed to overall radio, I control for the shares of the sub-prefecture with access to overall community radio, national and private radio. Standard errors are clustered at the prefecture level.
	
	
	
	\begin{figure}[H]
		\setstretch{1}\selectfont
		\begin{adjustwidth}{-0.5cm}{}
			\center	\caption{Relationship between learning about Ebola from the media, conditional on receiving some information about Ebola, and second-order beliefs (A) or own private health behavior (B) \label{fig:inseb_neigh}}
			\subcaption*{A. Belief about a socially-sanctioned behavior: \\
				\small{\textit{Belief that neighbors seek more health treatment}} } 
			\includegraphics[width=.7\columnwidth, trim=4 25 4 0,clip]{\figures/07_inseb/cstd_allmediatreatnowrel2025_\radio_3way.pdf} 
			\subcaption*{B. Private or not socially-costly action: \\
				\small{ \textit{Chlorine use for hand-washing/clothes}} }
			\includegraphics[width=.7\columnwidth, trim=4 25 4 0,clip]{\figures/07_inseb/cstd_allmediachlor2025_\radio_3way.pdf} \\ 
			%
			\vspace{-0.5cm}
		\end{adjustwidth}
		\begin{threeparttable}
			\begin{tablenotes}
				\item \footnotesize \qquad \qquad \qquad With 95\% confidence intervals; Robust standard errors clustered by prefecture. 
				\vspace{0.1cm}
				\item[] \footnotesize \textit{Notes}: Effects of learning about Ebola from the media, compared to other information sources (door-to-door campaigns, doctors, NGOs); conditional on above-median overall radio access and controlling for access to other radio stations. Outcomes are standardized variables. Seeking treatment takes three values (less/same/more than before). I plot coefficients on four different regressions based on equation (\ref{eq:info_impacts1}): (1) reduced form effect of percent of a sub-prefecture with access to a local community radio; (2) OLS effect of learning about Ebola from the media; (3) the effect of others in the sub-prefecture learning about Ebola from the media, except oneself; and (4) this effect depending on the percentage of others learning about Ebola from the media. All coefficients control for learning about Ebola from any source. The last two sets of regressions also control for others learning about Ebola from any source. I condition on above-median access to overall radio. I control for the shares of the sub-prefecture with access to overall community radio, national radio and private radio, for individual and household characteristics (age, gender, nationality, education, marital status, employment status, household size, number of household members working, number of students, number of children under five, drinking-water and lighting source, housing tenure and type, number of bedrooms, roof and floor materials, and a wealth index), and for log-population and geographic area at the sub-prefecture level. Tables A12-A14 show regression estimates (including 2SLS results). Data source: Post-Ebola survey (2015), Guinean National Institute of Statistics (INS).
			\end{tablenotes}
		\end{threeparttable}
	\end{figure}

	\subsubsection*{Results on the impact of local media on second-order beliefs vs private actions}

	The results indicate that learning about Ebola from the media has impacts beyond hearing about Ebola through other sources on socially-sanctioned behavior, but not on health behavior that is either private or not socially-sanctioned, Figure \ref{fig:inseb_neigh} (and Tables A13-A14).

	\paragraph{Second-order beliefs: neighbors seeking treatment.}
	
	People who learned about Ebola from the media are .3 of a standard deviation (SD) more likely to think their neighbors will seek treatment, conditional on having received any information about Ebola (OLS in Figure \ref{fig:inseb_neigh}; column 2 of Panel A of Table A13). The OLS coefficient implies that a 10 pp. increase in the fraction of others' hearing about Ebola from the media is associated with a .17 SD change in second order beliefs about others seeking treatment (Others in Figure \ref{fig:inseb_neigh}; column 2, Panel B of Table A13). This effect appears to be non-linear, with a greater fraction of people seeking treatment being associated with a larger effect, in line with a coordination mechanism (column 2, Table A15). The reduced form effect implies that 10 pp. greater local radio coverage is associated with a .08 SD rise in the belief that others will seek treatment. 
	
		

Turning to the IV strategy and examining the first stage results reveals that local community radio is a better predictor of others' hearing about Ebola in the media, than of having heard it oneself. The first stage is strong in both cases, when conditioning on comparable places |those with above-median access to radio (F-stat $>$ 13, columns 3/8 of Table A12; Kleibergen-Paap rk Wald F $>=$ 18, column 5 in both panels of Table A13). After conditioning on the leave-one-out share, however, the residual first-stage coefficient on the own dummy is roughly halved, while the coefficient on the leave-one-out share is essentially unchanged (column 5/10, Table A12). I therefore interpret the 2SLS estimates as identifying impacts of \textit{community-level} media exposure |i.e., the effects of knowing one's neighbors would have heard about Ebola from the media, rather than of receiving the relevant information| reinforcing the coordination argument. An increase by 10 pp. in the share of others who have heard about Ebola from the media leads to a .43 SD greater belief that others will seek treatment, larger than the corresponding OLS estimate (.17 SD per 10 pp.; equality of OLS and 2SLS is rejected at the 10\% level, p $=$ 0.07). Three aspects can explain the larger 2SLS estimate: measurement error in the share of individuals who heard about Ebola from the media, OLS capturing also the effects of selection of individuals who need more information from the media, and compliers being less informed than the average population.\footnote{If people whose exposure is shifted by radio coverage are those who otherwise would not have had access to good information about Ebola, then their gain from being exposed is plausibly larger than the average gain.}



\paragraph{Private actions: washing clothes/hands with chlorine.}

	On the other hand, I do not observe impacts of media on the use of chlorine for washing hands or clothes, conditional on receiving any information about Ebola (panel (b) of Figure \ref{fig:inseb_neigh}; Tables A14-A15). This is in line with the coordination effects of media being key for socially-sanctioned behaviors, but not necessarily for behaviors that are private or not socially costly. 
	
\paragraph{Local versus other radio.} The effects are specific to local radio. Local community radio coverage raises the belief that neighbors seek treatment (0.79 SD per unit of coverage, $p<0.05$) but, if anything, lowers chlorine use; reduced-form estimates for coverage by community radio sharing the respondent's language, by any community radio, or by national radio show no positive effect on either outcome (Figure A22; Table A16). This again points to local radio as the channel through which behavior changed.

\subsection{Evaluating trust as an explanation of the locality effect \label{sec:mech_trust}}	

Trust may well matter for how health information is absorbed in general, but does greater trust explain why \textit{local} radio outperforms non-local radio? If the effects of local radio were primarily due to it being a more trusted source |it should affect all types of behavior, whether socially sanctioned or private (see OA Section B.2.3). The previous Section \ref{sec:mech_beliefs} did not find impacts on private behavior; and I documented effects on a range of socially-sanctioned behaviors and second order beliefs (Sections \ref{sec:mech_social}- \ref{sec:mech_beliefs}), suggesting \textit{norm proximity} as a key channel. 
	
	To further examine the role of trust in explaining the impacts of local radio, as a mechanism, I test for heterogeneous treatment effects by prefecture-level trust in media, measured in the 2013 Afrobarometer, prior to the epidemic. If local radio mattered because its own community trusts it more than other stations, its impact should vary with pre-existing trust in media, provided that general trust in media also shifts how much people trust their local station relative to others (OA Section B.2.3). The evidence does not support this. The effect of local radio on Ebola cases does not vary with trust in media (Figure A20, panel b).\footnote{I construct a summary index statistic of three Afrobarometer 2013 survey questions: listening to news on the radio, believing that media is effective in revealing government corruption, and not agreeing with the statement that the media gives false information.} 
	
	
	I conduct an additional exercise to examine whether a stated preference for community radio (as the most reliable source of information, relative to other radio stations) is predictive of an earlier drop in infections. This data is only available in Afrobarometer 2017, after the end of the outbreak, so results are interpreted as suggestive. It is, however, a more direct test, since it measures trust in community radio relative to other stations. If the effects of local media were due to community radio being more trusted in places that have a local radio, we would see larger effects in places reporting greater preference for community radio. We see a large but not statistically-significant drop in infections in places that prefer community radio (Figure A21, panel b). But this pattern is equally consistent with people simply listening to community radio more where they prefer it. Since preferences are measured after the outbreak, they may also partly reflect the campaign's own success. Together with the evidence above |no impacts of local media on private behavior, and no heterogeneity by pre-existing trust in media| these results suggest that the effects of local media are not primarily driven by people trusting it more. \\
	
	Taken together, the evidence narrows down the channel through which local radio impacts behavior. Local radio affects socially-sanctioned behaviors but not private ones (Sections~\ref{sec:mech_social}-\ref{sec:mech_beliefs}), which rules out informational precision and trust as primary explanations of the locality effect, and its effect does not vary with pre-existing trust in media (Section~\ref{sec:mech_trust}), which weighs specifically against trust as the primary explanation of the locality effect. Trust may still facilitate the absorption of health information in general; the evidence indicates it does not explain the impacts of \textit{local} radio. The empirical results in Section~\ref{sec:empirical} show that effects come from \emph{geographically} local radio, rather than from radio sharing one's ethno-linguistic identity (Hypothesis 1b), and that a shared identity helps at first but high ethno-linguistic overlap is not required for local radio to work (Hypothesis 1c) |a pattern consistent with norm proximity, since norms can only be sanctioned by people who can observe them, which requires physical proximity. Norm proximity therefore emerges as the dominant channel, with informational precision or relevance playing a secondary role.

%

	\section{Conclusion \label{sec:conclusion}}
	
This paper provides first evidence on the impacts of local media in halting an epidemic outbreak, and novel evidence on the mechanism through which local media shifts norms: reaching the very community that observes and sanctions behavior, rather than a large, diffuse audience.

A public health campaign aired on community radio led to a significant and earlier reduction in Ebola infections in places with access to a \emph{local} community radio station, relative to neighboring places reached by the same radio but for which it was not local. A back-of-the-envelope calculation implies that 12--17\% of total Ebola cases in Guinea could have been prevented had all places with above-median access to community radio had their own local station. The drop in infections is accompanied by a coordinated change in socially-sanctioned health behaviors: opposition to public health measures declines, and treatment uptake rises, while private health behaviors do not respond. This coordination runs through the common-knowledge property of radio, illustrated by mobile-radio sensitization sharply reducing refusals of safe burials while door-to-door campaigns did not. The effects are driven by the locality of the information source, rather than by ethno-linguistic belonging, and operate through \emph{norm proximity}: local radio reaches the very community that can monitor and enforce the norm. 

The evidence points to a trade-off in the economics of media between the reach of communication and its ability to influence social norms. Media with greater coverage is typically expected to achieve greater coordination: a single broadcast reaches many listeners at once. But communication that reaches a large, diffuse audience updates beliefs about many, mostly distant, others; very local communication updates beliefs about fewer people, but people physically close by, who actually witness one's behavior. This paper shows that, for socially-sanctioned behaviors, observed in the physical space, locality outweighs reach.


These findings advance our understanding of how harmful community norms change. They can help design more effective interventions for improving health behavior in developing countries: information campaigns need to be distributed at the level at which norms are observed and sanctioned. Yet the implications of these findings extend well beyond epidemic response: the same coordination problem recurs wherever behavior is sanctioned by the immediate community rather than chosen privately —vaccine uptake, mask-wearing, gender norms, environmental compliance, or norms around corruption and harassment. Even in ethnically diverse settings, where ethno-linguistic identity is often assumed to define norms, the relevant group may be the local community that monitors and sanctions behavior.

From these findings follow promising directions for future work. A natural next step is examining whether norm change achieved locally then diffuses more broadly |whether local shifts extend into change at larger scale, or remain confined to the communities that experienced them. Another is whether social media's effects likewise operate more strongly through geographically close networks; and whether online behavior is governed more by the dispersed audiences that observe it, or still by those one will meet again in person.

	
	\bibliographystyle{\dirnew/bibtex/aea.bst}
	
	\singlespacing
	
	\bibliography{\dirnew/bibtex/bibliography_gn.bib}

@article{de2025stata,
	title={DID\_MULTIPLEGT\_DYN: Stata module to estimate event-study Difference-in-Difference (DID) estimators in designs with multiple groups and periods, with a potentially non-binary treatment that may increase or decrease multiple times},
	author={de Chaisemartin, Cl{\'e}ment and Ciccia, Diego and D'Haultfoeuille, Xavier and Knau, Felix and Mal{\'e}zieux, M{\'e}litine and Sow, Doulo},
	year={2025},
	publisher={Boston College Department of Economics},
	url={https://ideas.repec.org/c/boc/bocode/s459222.html},	
}

@article{de2024difference,
	title={Difference-in-differences estimators of intertemporal treatment effects},
	author={de Chaisemartin, Cl{\'e}ment and d'Haultf{\oe}uille, Xavier},
	journal={Review of Economics and Statistics},
	pages={1--45},
	year={2024},
	publisher={MIT Press One Rogers Street, Cambridge, MA 02142-1209, USA journals-info~…}
}

@article{lowes2021legacy,
	title={The legacy of colonial medicine in Central Africa},
	author={Lowes, Sara and Montero, Eduardo},
	journal={American Economic Review},
	volume={111},
	number={4},
	pages={1284--1314},
	year={2021},
	publisher={American Economic Association 2014 Broadway, Suite 305, Nashville, TN 37203}
}

@article{bursztyn2020extreme,
	title={From extreme to mainstream: The erosion of social norms},
	author={Bursztyn, Leonardo and Egorov, Georgy and Fiorin, Stefano},
	journal={American economic review},
	volume={110},
	number={11},
	pages={3522--3548},
	year={2020},
	publisher={American Economic Association 2014 Broadway, Suite 305, Nashville, TN 37203}
}

@inproceedings{de2019does,
	title={Does maternal education decrease female genital cutting?},
	author={De Cao, Elisabetta and La Mattina, Giulia},
	booktitle={AEA Papers and Proceedings},
	volume={109},
	pages={100--104},
	year={2019},
	organization={American Economic Association 2014 Broadway, Suite 305, Nashville, TN 37203}
}

@article{martinez2022vaccines,
	title={In vaccines we trust? The effects of the CIA’s vaccine ruse on immunization in Pakistan},
	author={Martinez-Bravo, Monica and Stegmann, Andreas},
	journal={Journal of the European Economic Association},
	volume={20},
	number={1},
	pages={150--186},
	year={2022},
	publisher={Oxford University Press}
}

@article{gulesci2025stepping,
	title={A stepping stone approach to norm transitions},
	author={Gulesci, Selim and Jindani, Sam and La Ferrara, Eliana and Smerdon, David and Sulaiman, Munshi and Young, Peyton},
	journal={American Economic Review},
	volume={115},
	number={7},
	pages={2237--2266},
	year={2025},
	publisher={American Economic Association 2014 Broadway, Suite 305, Nashville, TN 37203}
}

@article{corno2020age,
	title={Age of marriage, weather shocks, and the direction of marriage payments},
	author={Corno, Lucia and Hildebrandt, Nicole and Voena, Alessandra},
	journal={Econometrica},
	volume={88},
	number={3},
	pages={879--915},
	year={2020},
	publisher={Wiley Online Library}
}

@article{montero2022religious,
	title={Religious festivals and economic development: Evidence from the timing of mexican saint day festivals},
	author={Montero, Eduardo and Yang, Dean},
	journal={American Economic Review},
	volume={112},
	number={10},
	pages={3176--3214},
	year={2022},
	publisher={American Economic Association 2014 Broadway, Suite 305, Nashville, TN 37203}
}

@article{gadenne2014decentralization,
	title={Decentralization in developing economies},
	author={Gadenne, Lucie and Singhal, Monica},
	journal={Annu. Rev. Econ.},
	volume={6},
	number={1},
	pages={581--604},
	year={2014},
	publisher={Annual Reviews}
}

@article{allen2024correcting,
	title={Correcting Misperceptions about Support for Social Distancing to Combat COVID-19},
	author={Allen, James and Mahumane, Arlete and Riddell, James and Rosenblat, Tanya and Yang, Dean and Yu, Hang},
	journal={Economic Development and Cultural Change},
	volume={73},
	number={1},
	pages={221--242},
	year={2024},
	publisher={The University of Chicago Press Chicago, IL}
}

@article{yang2023knowledge,
	title={Knowledge, stigma, and HIV testing: An analysis of a widespread HIV/AIDS program},
	author={Yang, Dean and Allen, James and Mahumane, Arlete and Riddell, James and Yu, Hang},
	journal={Journal of Development Economics},
	volume={160},
	pages={102958},
	year={2023},
	publisher={Elsevier}
}

@article{lau2019social,
	title={Social norms and free-riding in influenza vaccine decisions in the UK: an online experiment},
	author={Lau, Krystal and Miraldo, Marisa and Galizzi, Matteo M and Hauck, Katharina},
	journal={The Lancet},
	volume={394},
	pages={S65},
	year={2019},
	publisher={Elsevier}
}

@article{martinez2021let,
	title={Let’s (not) get together! The role of social norms on social distancing during COVID-19},
	author={Martinez, Deborah and Parilli, Cristina and Scartascini, Carlos and Simpser, Alberto},
	journal={PloS one},
	volume={16},
	number={3},
	pages={e0247454},
	year={2021},
	publisher={Public Library of Science San Francisco, CA USA}
}

@article{dupas2024negligible,
	title={The Negligible Effect of Free Contraception on Fertility: Experimental Evidence from Burkina Faso},
	doi={10.1257/aer.20241305},
	author={Dupas, Pascaline and Jayachandran, Seema and Lleras-Muney, Adriana and Rossi, Pauline},
	journal={American Economic Review},
	volume={115},
	number={8},
	pages={2659--2688},
	year={2025}
}

@article{maffioli2021political,
	title={The political economy of health epidemics: Evidence from the Ebola outbreak},
	author={Maffioli, Elisa M},
	journal={Journal of Development Economics},
	volume={151},
	pages={102651},
	year={2021},
	publisher={Elsevier}
}

@article{pinna2022cable,
	title={Cable news and COVID-19 vaccine uptake},
	author={Pinna, Matteo and Picard, L{\'e}o and Goessmann, Christoph},
	journal={Scientific reports},
	volume={12},
	number={1},
	pages={16804},
	year={2022},
	publisher={Nature Publishing Group UK London}
}

@article{buhler2026couch,
	title={From Couch to Poll: Media Content and the Value of Local Information},
	author={B{\"u}hler, Mathias and Dickens, Andrew},
	journal={Journal of Public Economics},
	volume={256},
	year={2026}
}

@article{cornand2008optimal,
	title={Optimal degree of public information dissemination},
	author={Cornand, Camille and Heinemann, Frank},
	journal={The Economic Journal},
	volume={118},
	number={528},
	pages={718--742},
	year={2008},
	publisher={Oxford University Press Oxford, UK}
}

@article{kuznetsova2016value,
	title={The value of public information in a two-region model},
	author={Kuznetsova, Olga},
	journal={Higher School of Economics Research Paper No. WP BRP},
	volume={126},
	year={2016}
}

@article{shadmehr2020coordination,
	title={Coordination and social distancing: Inertia in the aggregate response to covid-19},
	author={Shadmehr, Mehdi and de Mesquita, Ethan Bueno},
	journal={University of Chicago, Becker Friedman Institute for Economics Working Paper},
	number={2020-53},
	year={2020}
}

@article{durante2012partisan,
	title={Partisan control, media bias, and viewer responses: Evidence from Berlusconi’s Italy},
	author={Durante, Ruben and Knight, Brian},
	journal={Journal of the European Economic Association},
	volume={10},
	number={3},
	pages={451--481},
	year={2012},
	publisher={Oxford University Press}
}

@article{enikolopov2011media,
	title={Media and political persuasion: Evidence from Russia},
	author={Enikolopov, Ruben and Petrova, Maria and Zhuravskaya, Ekaterina},
	journal={American Economic Review},
	volume={101},
	number={7},
	pages={3253--3285},
	year={2011},
	publisher={American Economic Association}
}

@article{grosfeld2024independent,
	title={Independent Media, Propaganda, and Religiosity: Evidence from Poland},
	author={Grosfeld, Irena and Madinier, Etienne and Sakalli, Seyhun Orcan and Zhuravskaya, Ekaterina},
	journal={American Economic Journal: Applied Economics},
	volume={16},
	number={4},
	pages={361--403},
	year={2024}
}

@article{woodcock2021role,
	title={The role of conformity in mask-wearing during COVID-19},
	author={Woodcock, Anna and Schultz, P Wesley},
	journal={Plos one},
	volume={16},
	number={12},
	pages={e0261321},
	year={2021},
	publisher={Public Library of Science San Francisco, CA USA}
}

@article{geanakoplos1992common,
	title={Common knowledge},
	author={Geanakoplos, John},
	journal={Journal of Economic Perspectives},
	volume={6},
	number={4},
	pages={53--82},
	year={1992}
}

@book{blakeman2011advertising,
	title={Advertising campaign design: Just the essentials},
	author={Blakeman, Robyn},
	year={2011},
	publisher={Routledge}
}

@article{wakefield2011effects,
	title={Effects of mass media campaign exposure intensity and durability on quit attempts in a population-based cohort study},
	author={Wakefield, MA and Spittal, MJ and Yong, HH and Durkin, SJ and Borland, R},
	journal={Health Education Research},
	volume={26},
	number={6},
	pages={988--997},
	year={2011},
	publisher={Oxford University Press}
}

@article{breza2021effects,
	title={Effects of a large-scale social media advertising campaign on holiday travel and COVID-19 infections: a cluster randomized controlled trial},
	author={Breza, Emily and Stanford, Fatima Cody and Alsan, Marcella and Alsan, Burak and Banerjee, Abhijit and Chandrasekhar, Arun G and Eichmeyer, Sarah and Glushko, Traci and Goldsmith-Pinkham, Paul and Holland, Kelly and others},
	journal={Nature medicine},
	volume={27},
	number={9},
	pages={1622--1628},
	year={2021},
	publisher={Nature Publishing Group}
}

@book{hornik2002public,
	title={Public health communication: Evidence for behavior change},
	author={Hornik, Robert},
	year={2002},
	publisher={Routledge}
}

@article{flay1987mass,
	title={Mass media and smoking cessation: a critical review.},
	author={Flay, Brian R},
	journal={American Journal of Public Health},
	volume={77},
	number={2},
	pages={153--160},
	year={1987},
	publisher={American Public Health Association}
}

@article {nunnpuga2012,
	title = {Ruggedness: The Blessing of Bad Geography in Africa},
	journal = {Review of Economics and Statistics},
	volume = { 94},
	number = {1},
	year = {2012},
	pages = { 20-36},
	author = {Nathan Nunn and Diego Puga}
}

@techreport{glennerster2026media,
	title={Media, social pressure, and combating misinformation: Experimental evidence on mass media and contraception use in Burkina Faso},
	author={Glennerster, Rachel and Murray, Joanna and Pouliquen, Victor},
	year={2026},
	institution={Working paper}
}

@article{armand2020reach,
	title={The reach of radio: Ending civil conflict through rebel demobilization},
	author={Armand, Alex and Atwell, Paul and Gomes, Joseph F},
	journal={American economic review},
	volume={110},
	number={5},
	pages={1395--1429},
	year={2020}
}

@article{brodeur2021literature,
	title={A literature review of the economics of COVID-19},
	author={Brodeur, Abel and Gray, David and Islam, Anik and Bhuiyan, Suraiya},
	journal={Journal of Economic Surveys},
	volume={35},
	number={4},
	pages={1007--1044},
	year={2021},
	publisher={Wiley Online Library}
}

@article{desmet2017culture,
	title={Culture, ethnicity, and diversity},
	author={Desmet, Klaus and Ortu{\~n}o-Ort{\'\i}n, Ignacio and Wacziarg, Romain},
	journal={American Economic Review},
	volume={107},
	number={9},
	pages={2479--2513},
	year={2017}
}

@article{alesina2002trusts,
	title={Who trusts others?},
	author={Alesina, Alberto and La Ferrara, Eliana},
	journal={Journal of public economics},
	volume={85},
	number={2},
	pages={207--234},
	year={2002},
	publisher={Elsevier}
}

@article{alesina2005ethnic,
	title={Ethnic diversity and economic performance},
	author={Alesina, Alberto and La Ferrara, Eliana},
	journal={Journal of economic literature},
	volume={43},
	number={3},
	pages={762--800},
	year={2005}
}

@article{michalopoulos2016long,
	title={The long-run effects of the scramble for Africa},
	author={Michalopoulos, Stelios and Papaioannou, Elias},
	journal={American Economic Review},
	volume={106},
	number={7},
	pages={1802--48},
	year={2016}
}

@article{oberholzer2009media,
	title={Media markets and localism: Does local news en Espanol boost Hispanic voter turnout?},
	author={Oberholzer-Gee, Felix and Waldfogel, Joel},
	journal={American Economic Review},
	volume={99},
	number={5},
	pages={2120--28},
	year={2009}
}

@article{karing2024social,
	title={Social signaling and childhood immunization: A field experiment in Sierra Leone},
	author={Karing, Anne},
	journal={The Quarterly Journal of Economics},
	volume={139},
	number={4},
	pages={2083--2133},
	year={2024},
	publisher={Oxford University Press}
}

@article{stromberg2015media,
	title={Media and politics},
	author={Str{\"o}mberg, David},
	journal={Annual Review of Economics},
	volume={7},
	number={1},
	pages={173--205},
	year={2015},
	publisher={Annual Reviews}
}

@article{james2020,
	author= {James, Peter Bai and Wardle, Jonathan and Steel, Amie and Adams, Jon},
	journal = {BMC Public Health},
	number = {1},
	year = {2020},
	pages = {182},
	title = {An assessment of Ebola-related stigma and its association with informal healthcare utilisation among Ebola survivors in Sierra Leone: a cross-sectional study},
}

@article{richards2015social,
  title={Social pathways for Ebola virus disease in rural Sierra Leone, and some implications for containment},
  author={Richards, Paul and Amara, Joseph and Ferme, Mariane C and Kamara, Prince and Mokuwa, Esther and Sheriff, Amara Idara and Suluku, Roland and Voors, Maarten},
  journal={PLoS Negl Trop Dis},
  volume={9},
  number={4},
  year={2015},
  publisher={Public Library of Science}
}

@article{garske2017heterogeneities,
  title={Heterogeneities in the case fatality ratio in the West African Ebola outbreak 2013--2016},
  author={Garske, Tini and Cori, Anne and Ariyarajah, Archchun and Blake, Isobel and Dorigatti, Ilaria and Eckmanns, Tim and Fraser, Christophe and Hinsley, Wes and Jombart, Thibaut and Mills, Harriet},
  journal={Philosophical Transactions of the Royal Society: Biological Sciences},
  volume={372},
  number={1721},
  year={2017},
  publisher={The Royal Society}
}

@article{wakefield2010use,
	title={Use of mass media campaigns to change health behaviour},
	doi={10.1016/S0140-6736(10)60809-4},
	author={Wakefield, Melanie A and Loken, Barbara and Hornik, Robert C},
	journal={The Lancet},
	volume={376},
	number={9748},
	pages={1261--1271},
	year={2010},
	publisher={Elsevier}
}

@book{katz1966personal,
	title={Personal Influence, The part played by people in the flow of mass communications},
	author={Katz, Elihu and Lazarsfeld, Paul Felix},
	year={1966},
	publisher={Transaction publishers}
}

@article{van2020traditional,
	title={Traditional leaders and the 2014--2015 Ebola epidemic},
	author={Van der Windt, Peter and Voors, Maarten},
	journal={The Journal of Politics},
	volume={82},
	number={4},
	pages={1607--1611},
	year={2020},
	publisher={The University of Chicago Press Chicago, IL}
}

@article{christensen2021healthcare,
  title={Building Resilient Health Systems: Experimental Evidence from Sierra Leone and the 2014 Ebola Outbreak},
  author={Christensen, Darin and Dube, Oeindrila and Haushofer, Johannes and Siddiqi, Bilal and Voors, Maarten},
  journal={The Quarterly Journal of Economics},
  volume={136},
  number={2},
  pages={1145--1198},
  year={2021}
}

@article{gonzalez2025epidemics,
	title={Epidemics and conflict: Evidence from the Ebola outbreak in Western Africa},
	author={Gonz\'{a}lez-Torres, Ada and Esposito, Elena},
	journal={Available at SSRN 3544606},
	year={2025}
}

@article{banerjee2024less,
	title={When Less Is More: Experimental Evidence on Information Delivery During India’s Demonetisation},
	author={Banerjee, Abhijit and Breza, Emily and Chandrasekhar, Arun G and Golub, Benjamin},
	journal={Review of Economic Studies},
	volume={91},
	number={4},
	pages={1884--1922},
	year={2024},
	publisher={Oxford University Press UK}
}

@article{banerjee2010improving,
	title={Improving immunisation coverage in rural India: clustered randomised controlled evaluation of immunisation campaigns with and without incentives},
	author={Banerjee, Abhijit Vinayak and Duflo, Esther and Glennerster, Rachel and Kothari, Dhruva},
	journal={Bmj},
	volume={340},
	year={2010},
	publisher={British Medical Journal Publishing Group}
}

@techreport{banerjee2019entertaining,
	title={The entertaining way to behavioral change: Fighting HIV with MTV},
	author={Banerjee, Abhijit and La Ferrara, Eliana and Orozco-Olvera, Victor H},
	year={2019},
	institution={National Bureau of Economic Research}
}

@article{green2020countering,
	title={Countering violence against women by encouraging disclosure: A mass media experiment in rural Uganda},
	author={Green, Donald P and Wilke, Anna M and Cooper, Jasper},
	journal={Comparative Political Studies},
	volume={53},
	number={14},
	pages={2283--2320},
	year={2020},
	publisher={SAGE Publications Sage CA: Los Angeles, CA}
}

@article{adena2015radio,
  title={Radio and the Rise of the Nazis in Prewar Germany},
  author={Adena, Maja and Enikolopov, Ruben and Petrova, Maria and Santarosa, Veronica and Zhuravskaya, Ekaterina},
  journal={The Quarterly Journal of Economics},
  volume={130},
  number={4},
  pages={1885--1939},
  year={2015},
  publisher={MIT Press}
}

@article{adda2016economic,
  title={Economic activity and the spread of viral diseases: Evidence from high frequency data},
  author={Adda, J{\'e}r{\^o}me},
  journal={The Quarterly Journal of Economics},
  year={2016}
}

@book{angrist2008mostly,
  title={Mostly harmless econometrics: An empiricist's companion},
  author={Angrist, Joshua D and Pischke, J{\"o}rn-Steffen},
  year={2008},
  publisher={Princeton university press}
}

@incollection{bisin2011economics,
  title={The economics of cultural transmission and socialization},
  author={Bisin, Alberto and Verdier, Thierry},
  booktitle={Handbook of social economics},
  volume={1},
  pages={339--416},
  year={2011},
  publisher={Elsevier}
}

@article{bursztyn2018misperceived,
  title={Misperceived social norms: Female labor force participation in Saudi Arabia},
  author={Bursztyn, Leonardo and Gonz{\'a}lez, Alessandra L and Yanagizawa-Drott, David},
  journal={American Economic Review},
  volume={110},
  number={10},
  pages={2997--3029},
  year={2020}
}

@article{durante2018attack,
	title={Attack when the world is not watching? US news and the Israeli-Palestinian conflict},
	author={Durante, Ruben and Zhuravskaya, Ekaterina},
	journal={Journal of Political Economy},
	volume={126},
	number={3},
	pages={1085--1133},
	year={2018},
	publisher={University of Chicago Press Chicago, IL}
}

@article{cantoni2017protests,
  title={Are Protests Games of Strategic Complements or Substitutes? Experimental Evidence from Hong Kong's Democracy Movement},
  author={Cantoni, Davide and Yang, David Y and Yuchtman, Noam and Zhang, Y Jane},
  journal={The Quarterly Journal of Economics},
  volume={134},
  number={2},
  pages={1007--1062},
  year={2019}
}

@article{dellavigna2007fox,
  title={The Fox News effect: Media bias and voting},
  author={DellaVigna, Stefano and Kaplan, Ethan},
  journal={The Quarterly Journal of Economics},
  volume={122},
  number={3},
  pages={1187--1234},
  year={2007},
  publisher={MIT Press}
}

@article{dupas2011teenagers,
	title={Do teenagers respond to HIV risk information? Evidence from a field experiment in Kenya},
	doi={10.1257/app.3.1.1},
	author={Dupas, Pascaline},
	journal={American Economic Journal: Applied Economics},
	volume={3},
	number={1},
	pages={1--34},
	year={2011},
	publisher={American Economic Association}
}

@article{fearon1996explaining,
  title={Explaining interethnic cooperation},
  author={Fearon, James D and Laitin, David D},
  journal={American political science review},
  volume={90},
  number={4},
  pages={715--735},
  year={1996},
  publisher={Cambridge University Press}
}

@article{fearon2003ethnicity,
  title={Ethnicity, insurgency, and civil war},
  author={Fearon, James D and Laitin, David D},
  journal={American political science review},
  volume={97},
  number={1},
  pages={75--90},
  year={2003},
  publisher={Cambridge University Press}
}

@article{wang2020black,
	title={Waves of Empowerment: Black Radio and the Civil Rights Movement},
	author={Wang, Tianyi},
	journal={Working Paper},
	year={2020}
}

@article{gulzar2025good,
	title={Good politicians: Experimental evidence on motivations for political candidacy and government performance},
	author={Gulzar, Saad and Khan, Muhammad Yasir},
	journal={Review of Economic Studies},
	volume={92},
	number={1},
	pages={339--364},
	year={2025},
	publisher={Oxford University Press UK}
}

@article{ferrara2012soap,
  title={Soap Operas and Fertility: Evidence from Brazil},
  author={La Ferrara, Eliana and Chong, Alberto and Duryea, Suzanne},
  journal={American Economic Journal: Applied Economics},
  volume={4},
  number={4},
  pages={1--31},
  year={2012}
}

@article{ferraz2008exposing,
  title={Exposing corrupt politicians: the effects of Brazil's publicly released audits on electoral outcomes},
  author={Ferraz, Claudio and Finan, Frederico},
  journal={The Quarterly Journal of Economics},
  volume={123},
  number={2},
  pages={703--745},
  year={2008},
  publisher={MIT Press}
}

@article{galiani2016promoting,
  title={Promoting Handwashing Behavior: The Effects of Large-scale Community and School-level Interventions},
  doi={10.1002/hec.3273},
  author={Galiani, Sebastian and Gertler, Paul and Ajzenman, Nicolas and Orsola-Vidal, Alexandra},
  journal={Health economics},
  volume={25},
  number={12},
  pages={1545--1559},
  year={2016},
  publisher={Wiley Online Library}
}

@article{gershman2017long,
  title={Long-run development and the new cultural economics},
  author={Gershman, Boris},
  journal={Demographic Change and Long-Run Development},
  year={2017},
  publisher={MIT Press}
}

@article{jensen2009power,
  title={The power of TV: Cable television and women's status in India},
  author={Jensen, Robert and Oster, Emily},
  journal={The Quarterly Journal of Economics},
  volume={124},
  number={3},
  pages={1057--1094},
  year={2009},
  publisher={MIT Press}
}

@book{keefer2011mass,
  title={Mass media and public services: The effects of radio access on public education in Benin},
  author={Keefer, Philip and Khemani, Stuti},
  year={2011},
  publisher={The World Bank}
}

@article{keefer2014mass,
  title={Mass media and public education: The effects of access to community radio in Benin},
  author={Keefer, Philip and Khemani, Stuti},
  journal={Journal of Development Economics},
  volume={109},
  pages={57--72},
  year={2014},
  publisher={Elsevier}
}

@article{besley2002political,
  title={The political economy of government responsiveness: Theory and evidence from India},
  author={Besley, Timothy and Burgess, Robin},
  journal={The Quarterly Journal of Economics},
  volume={117},
  number={4},
  pages={1415--1451},
  year={2002},
  publisher={MIT Press}
}

@article{duflo2015education,
  title={Education, HIV, and early fertility: Experimental evidence from Kenya},
  doi={10.1257/aer.20121607},
  author={Duflo, Esther and Dupas, Pascaline and Kremer, Michael},
  journal={American Economic Review},
  volume={105},
  number={9},
  pages={2757--2797},
  year={2015},
  publisher={American Economic Association}
}

@article{eisensee2007news,
  title={News droughts, news floods, and US disaster relief},
  author={Eisensee, Thomas and Str{\"o}mberg, David},
  journal={The Quarterly Journal of Economics},
  volume={122},
  number={2},
  pages={693--728},
  year={2007},
  publisher={MIT Press}
}

@article{nickell1981biases,
  title={Biases in dynamic models with fixed effects},
  author={Nickell, Stephen},
  journal={Econometrica: Journal of the Econometric Society},
  year={1981}
}

@article{kremer2007illusion,
  title={The illusion of sustainability},
  author={Kremer, Michael and Miguel, Edward},
  journal={The Quarterly Journal of Economics},
  volume={122},
  number={3},
  pages={1007--1065},
  year={2007},
  publisher={MIT Press}
}

@article{morris2001global,
  title={Global games: Theory and applications},
  author={Morris, Stephen and Shin, Hyun Song},
  year={2001}
}

@article{morris2002social,
	title={Social value of public information},
	author={Morris, Stephen and Shin, Hyun Song},
	journal={American Economic Review},
	volume={92},
	number={5},
	pages={1521--1534},
	year={2002},
	publisher={American Economic Association}
}

@article{olken2009television,
  title={Do television and radio destroy social capital? Evidence from Indonesian villages},
  author={Olken, Benjamin A},
  journal={American Economic Journal: Applied Economics},
  volume={1},
  number={4},
  pages={1--33},
  year={2009},
  publisher={American Economic Association}
}

@book{richards2016ebola,
  title={Ebola: how a people's science helped end an epidemic},
  author={Richards, Paul},
  year={2016},
  publisher={Zed Books Ltd.}
}

@article{campante2024virus,
	title={The Virus of Fear: The Political Impact of Ebola in the United States},
	author={Campante, Filipe and Depetris-Chauvin, Emilio and Durante, Ruben},
	journal={American Economic Journal: Applied Economics},
	volume={16},
	number={1},
	pages={480--509},
	year={2024},
	publisher={American Economic Association 2014 Broadway, Suite 305, Nashville, TN 37203-2425}
}

@article{stromberg2004radio,
  title={Radio's impact on public spending},
  author={Str{\"o}mberg, David},
  journal={The Quarterly Journal of Economics},
  volume={119},
  number={1},
  pages={189--221},
  year={2004},
  publisher={MIT Press}
}

@article{fang2016transmission,
  title={Transmission dynamics of Ebola virus disease and intervention effectiveness in Sierra Leone},
  author={Fang, Li-Qun and Yang, Yang and Jiang, Jia-Fu and Yao, Hong-Wu and Kargbo, David and Li, Xin-Lou and Jiang, Bao-Gui and Kargbo, Brima and Tong, Yi-Gang and Wang, Ya-Wei and others},
  journal={Proceedings of the National Academy of Sciences},
  volume={113},
  number={16},
  pages={4488--4493},
  year={2016},
  publisher={National Acad Sciences}
}

@article{banerjee2024messages,
  title={Can a Trusted Messenger Change Behavior When Information Is Plentiful? Evidence from the First Months of the COVID-19 Pandemic in West Bengal},
  doi={10.1162/rest_a_01500},
  author={Banerjee, Abhijit and Alsan, Marcella and Breza, Emily and Chandrasekhar, Arun G and Chowdhury, Abhijit and Duflo, Esther and Goldsmith-Pinkham, Paul and Olken, Benjamin A},
  journal={Review of Economics and Statistics},
  year={2024},
  publisher={MIT Press}
}

@article{card2011family,
  title={Family violence and football: The effect of unexpected emotional cues on violent behavior},
  author={Card, David and Dahl, Gordon B},
  journal={The Quarterly Journal of Economics},
  volume={126},
  number={1},
  pages={103--143},
  year={2011},
  publisher={MIT Press}
}

@article{dellavigna2014cross,
  title={Cross-border effects of foreign media: Serbian radio and nationalism in Croatia},
  author={DellaVigna, Stefano and Enikolopov, Ruben and Mironova, Vera and Petrova, Maria and Zhuravskaya, Ekaterina},
  journal={American Economic Journal: Applied Economics},
  volume={6},
  number={3},
  pages={103--132},
  year={2014}
}

@incollection{dellavigna2015economic,
  title={Economic and social impacts of the media},
  author={DellaVigna, Stefano and La Ferrara, Eliana},
  booktitle={Handbook of Media Economics},
  volume={1A},
  chapter={19},
  pages={723--768},
  publisher={Elsevier},
  year={2015}
}

@article{lekone2006statistical,
  title={Statistical inference in a stochastic epidemic SEIR model with control intervention: Ebola as a case study},
  author={Lekone, Phenyo E and Finkenst{\"a}dt, B{\"a}rbel F},
  journal={Biometrics},
  volume={62},
  number={4},
  pages={1170--1177},
  year={2006},
  publisher={Wiley Online Library}
}

@article{funk2009spread,
  title={The spread of awareness and its impact on epidemic outbreaks},
  author={Funk, Sebastian and Gilad, Erez and Watkins, Chris and Jansen, Vincent AA},
  journal={Proceedings of the National Academy of Sciences},
  volume={106},
  number={16},
  pages={6872--6877},
  year={2009},
  publisher={National Acad Sciences}
}

@article{kiss2010impact,
  title={The impact of information transmission on epidemic outbreaks},
  author={Kiss, Istvan Z and Cassell, Jackie and Recker, Mario and Simon, P{\'e}ter L},
  journal={Mathematical biosciences},
  volume={225},
  number={1},
  pages={1--10},
  year={2010},
  publisher={Elsevier}
}

@article{paluck2009deference,
  title={Deference, dissent, and dispute resolution: An experimental intervention using mass media to change norms and behavior in Rwanda},
  author={Paluck, Elizabeth Levy and Green, Donald P},
  journal={American Political Science Review},
  volume={103},
  number={4},
  pages={622--644},
  year={2009},
  publisher={Cambridge University Press}
}

@article{blouin2019erasing,
	title={Erasing ethnicity? Propaganda, nation building, and identity in Rwanda},
	author={Blouin, Arthur and Mukand, Sharun W},
	journal={Journal of Political Economy},
	volume={127},
	number={3},
	pages={1008--1062},
	year={2019},
	publisher={The University of Chicago Press Chicago, IL}
}

@article{yanagizawa2014propaganda,
  title={Propaganda and conflict: Evidence from the Rwandan genocide},
  author={Yanagizawa-Drott, David},
  journal={The Quarterly Journal of Economics},
  volume={129},
  number={4},
  pages={1947--1994},
  year={2014},
  publisher={Oxford University Press}
}

@book{mcgovern2012unmasking,
  title={Unmasking the state: making Guinea modern},
  author={McGovern, Mike},
  year={2012},
  publisher={University of Chicago Press}
}

@article{goerg2011couper,
  title={Couper la Guin{\'e}e en quatre ou comment la colonisation a imagin{\'e} l'Afrique},
  author={Goerg, Odile},
  journal={Vingti{\`e}me Si{\`e}cle. Revue d'histoire},
  number={3},
  pages={73--88},
  year={2011},
  publisher={Presses de Sciences Po}
}

\end{document}